\documentclass[12pt,preprint]{aastex}
\input colordvi
\newcommand{\ltwid}{\mathrel{\raise.3ex\hbox{$<$\kern-.75em\lower1ex\hbox{$\sim$
}}}}
\newcommand{\gtwid}{\mathrel{\raise.3ex\hbox{$>$\kern-.75em\lower1ex\hbox{$\sim$
}}}}
\newcommand{\bd}{\begin{description}}
\newcommand{\ed}{\end{description}}
\usepackage[utopia]{mathdesign}
\usepackage{amsmath}
\newcommand{\s}{\scriptscriptstyle}
\newcommand{\bs}{\boldsymbol}

\begin{document}
\begin{center} Effect of core--mantle and tidal torques on Mercury's
spin axis orientation\\
Stanton J. Peale$^a$, Jean-Luc Margot$^{b,c}$, Steven A. Hauck, II$^d$,
Sean C. Solomon$^{e,f}$\\ 
\end{center}
$^a$Department of Physics, University of California, Santa Barbara,
CA 93106.\\
$^b$Department of Earth, Planetary, and Space Sciences, University of
California, Los Angeles CA 90095.\\ 
$^c$Department of Physics and Astronomy, University of California, Los
Angeles, CA 90095.\\
$^d$Department of Earth, Environmental, and Planetary Sciences, Case
Western Reserve University, Cleveland, OH 44106.\\
$^e$Department of Terrestrial Magnetism, Carnegie Institution of
Washington, Washington, DC 20015.\\
$^f$Lamont-Doherty Earth Observatory, Columbia University, Palisades,
NY 10964.\\ 
\vspace{.1in}
\begin{center}
{\bf Abstract}
\end{center}
The rotational evolution of Mercury's mantle plus crust and its
core under conservative and dissipative torques is important for
understanding the planet's spin state. Dissipation results from tidal
torques and viscous, magnetic, and topographic torques
contributed by interactions between the liquid core and solid mantle.
For a spherically symmetric core--mantle boundary (CMB), the
system goes to an equilibrium state wherein the spin axes of the mantle
and core are fixed in the frame precessing with the orbit, and in which
the mantle and core are differentially rotating. This equilibrium
exhibits a mantle spin axis that is offset from the Cassini state by
larger amounts for weaker core--mantle coupling for all three
dissipative core--mantle coupling 
mechanisms, and the spin axis of the core is separated farther from that
of the mantle, leading to larger differential rotation.  Relatively
strong core--mantle coupling is necessary to bring the mantle spin axis
to a position within the uncertainty in its observed
position, which is close to the Cassini state defined for a completely
solid Mercury. Strong core--mantle coupling means that Mercury's
response is closer to that of a solid planet. Measured or
inferred  values of   
parameters in all three core--mantle coupling mechanisms for a
spherically symmetric CMB  appear not to accomplish this
requirement. For a hydrostatic ellipsoidal CMB, pressure coupling
dominates the dissipative effects on the mantle and core positions,
and dissipation with pressure coupling brings the mantle spin
solidly to the Cassini state. The core spin goes to a position
displaced from that of the mantle by about 3.55 arcmin nearly in the
plane containing the Cassini state.  The core spin lags the precessing
plane containing the Cassini state by an increasing angle as the core
viscosity is increased.  With the maximum viscosity considered of
$\nu\sim 15.0\,{\rm cm^2/s}$ if the coupling is by the circulation
through an Ekman boundary layer or $\nu\sim 8.75\times 10^5\,{\rm
cm^2/s}$ for purely viscous coupling, the core spin lags the
precessing Cassini plane by 23 arcsec, whereas the mantle spin lags by
only 0.055 arcsec. Larger, non hydrostatic values of the CMB
ellipticity also result in the mantle spin at the Cassini state, but the
core spin is moved closer to the mantle spin.  Current measurement
uncertainties preclude using the mantle offset to constrain the
internal core viscosity.
\newpage

\section{Introduction\label{sec:introduction}}
Mercury is in a stable spin--orbit resonance in which the rotational
angular velocity is precisely 1.5 times the mean orbital motion 
(Pettengill and Dyce, 1965; Colombo and Shapiro, 1966). This rotation state 
is a natural outcome of tidal evolution (Goldreich and Peale,
1966; Correia and Laskar, 2004, 2009). In addition, the same
tidal evolution brings Mercury to Cassini state 1, wherein Mercury's
spin axis remains coplanar with the orbit normal and Laplace plane
normal as the spin vector and orbit normal precess
around the latter with a $\sim 300,000$ yr period (Colombo 1966, Peale,
1969, 1974). That Mercury is very close to this state has been
verified with radar observations, which give an obliquity of $2.04\pm
0.08$ arcmin (Margot {\it et al.}, 2007, 2012). The most recent
observations show that the best-fit solution is offset from the
Cassini state by a few arcseconds, but the uncertainty at one standard
deviation includes the Cassini state.

This paper is an investigation of the possible displacement of the
spin axis from the Cassini state from dissipative processes and the
consequences of pressure coupling. In Section \ref{sec:equations} we 
develop the equations for the rotational motion of both the core and
mantle plus crust from conservative and dissipative torques.  The
latter include the tidal torque and the torques due to viscous,
magnetic, and topographical coupling between the core and the mantle
plus crust for a spherically 
symmetric core--mantle boundary (CMB). Gravitational and rotational
distortions of the CMB lead to pressure torques that dominate all the
dissipative torques. Results are given in Section \ref{sec:results},
where we show that the tidal offset of the mantle spin axis from the
Cassini state is immeasurably small, but the offset 
due to the core--mantle interactions can be quite large, and
weaker core--mantle coupling leads to larger offsets.  The core--mantle
dissipative coupling must be relatively strong to bring the mantle
spin--axis to within the  uncertainty of its observed
location. The failure of viscous, magnetic and topographic mechanisms,
which dominate the tidal mechanism,  to
bring the spin axis near its observed position    
for measured or likely values of the parameters is compensated by the
pressure coupling between the core and mantle for both hydrostatic and
non--hydrostatic ellipsoidal CMB, which we examine in Section
\ref{sec:pressure}.  

We maintain the current orbital configurations throughout the
calculations even though the dissipative time scales are long enough
for significant changes to occur.  This assumption is justified
because the spin axis will follow the Cassini state as the latter's
position changes during the slow changes in the solar system
configuration because of adiabatic invariance of the solid angle swept
out by the spin axis as it precesses around the Cassini state. The spin
axis remains within 1 arcsec of the Cassini state position through
both long-period and short-period changes in the state position
(Peale, 2006).  We are interested only in the final equilibrium
positions of the core and mantle spins in the current orbit frame of
reference, and these positions will be the same if the evolution takes
place with the current, fixed orbital and solar system  parameters or
if these parameters are allowed to evolve during the evolution to the
current state.  

\section{Equations of variation\label{sec:equations}}
\begin{figure}[h]
\begin{center} 
\epsscale{.7}
\plotone{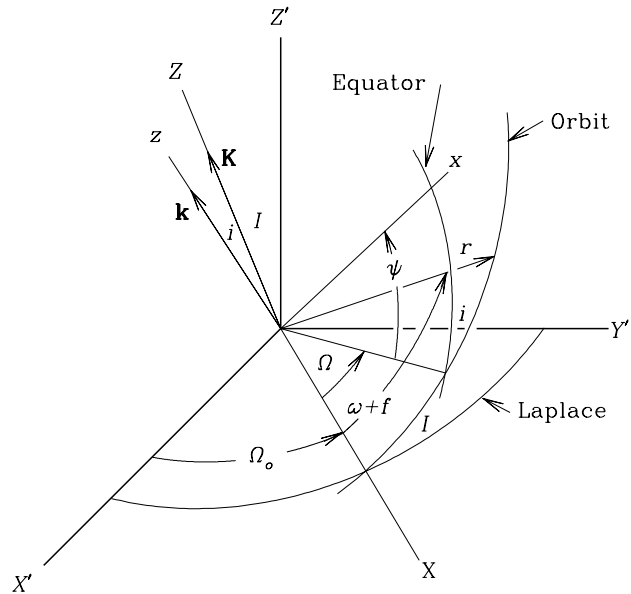} 
\caption{Coordinate systems and relevant angles. The angles orienting
mantle or core relative to the $XYZ$ orbit system will have subscripts
$m$ or $c$, respectively.  
\label{fig:coordsystems}}
\end{center}
\end{figure}

The coordinate systems and angles for the equations that govern the
rotational motion of Mercury are shown in Fig.
\ref{fig:coordsystems}, where $X^\prime, Y^\prime, Z^\prime$ are
quasi-inertial axes with the $X^\prime Y^{\prime}$ plane being the
Laplace plane on which Mercury's orbit precesses at nearly a constant
inclination $I$ and nearly constant angular velocity $\boldsymbol\mu$. The
$XYZ$ orbit system has the $X$ axis along the ascending node of the
orbit plane on the Laplace plane, and the $XY$ plane is the orbit
plane. The $xyz$ system is fixed in the body, with $z$ along the spin
axis and $x$ along the the axis of minimum moment of inertia in the
equator plane. The Euler angles orienting the $xyz$ system relative to
the $XYZ$ system are $\Omega,i,\psi$, where $\Omega$ is
the longitude of the ascending node of the equator plane on the $XY$
orbit plane measured from the $X$ axis, $i$ is the inclination of the
equator plane to the orbit plane, and $\psi$ is the angle from the
ascending node of the equator on the orbit plane to the $x$ axis of
minimum moment of inertia. The three Euler angles will have subscripts
$m$ or $f$ to designate mantle or fluid core, respectively. Angle 
$I$ is the inclination of the orbit plane to the Laplace plane,
$\Omega_o$ is the longitude of the ascending node of the orbit plane on
the Laplace plane, $\omega$ is the argument of perihelion, $f$ is the true
anomaly of the Sun, and $r$ is the distance from Mercury to
the Sun.

We assume principal axis rotation 
throughout.  The angular momentum of the mantle plus crust is ${\bf
L_m}=C_m\dot\psi_m 
{\bf k}_m=C_m\bs{\dot{\psi}_m}$, where $C_m$ is the moment of inertia of the
mantle plus crust about the spin 
axis, $\dot\psi_m$ is the angular velocity of the mantle, and 
${\bf k}_m=\sin{i_m}\sin{\Omega_m}{\bf I}-\sin{i_m}\cos{\Omega_m}{\bf
J}+\cos{i_m}{\bf K}$ is a unit vector along the spin axis. $\bf I,J,K$
are unit vectors along the $X,Y,Z$ axes, respectively. With $d{\bf L}_m/dt=
C_m(d\dot\psi_m/dt){\bf k}_m+C_m\dot\psi_m(d{\bf k}_m/dt)$, we can
write
\begin{eqnarray}
\frac{1}{C_m}\frac{dL_{m{\s X}}}{dt}&=&\frac{d\dot\psi_m}{dt}
\sin{i_m}\sin{\Omega_m} +\dot\psi_m\left[\cos{i_m}\sin{\Omega_m}
\frac{di_m}{dt}+\sin{i_m}\cos{\Omega_m}\frac{d\Omega_m}{dt}\right]\nonumber\\
\frac{1}{C_m}\frac{dL_{m{\s Y}}}{dt}&=&-\frac{d\dot\psi_m}{dt}
\sin{i_m}\cos{\Omega_m} +\dot\psi_m\left[-\cos{i_m}\cos{\Omega_m}
\frac{di_m}{dt}+\sin{i_m}\sin{\Omega_m}\frac{d\Omega_m}{dt}\right]\nonumber\\
\frac{1}{C_m}\frac{dL_{m{\s Z}}}{dt}&=&\frac{d\dot\psi_m}{dt}\cos{i_m}-
\dot\psi_m\sin{i_m}\frac{di_m}{dt}\label{eq:dLdt}
\end{eqnarray}
for the variations of the three components of angular momentum 
relative to the orbit system of coordinates, which system is readily
observable. 

The total torque on Mercury's mantle plus crust,  $\langle {\bf
T_m}\rangle=\langle{\bf T}_{body}\rangle+\langle{\bf T}_{tide}\rangle
+\langle{\bf T}_{f-m}\rangle$, is the sum of the conservative
gravitational torque, the tidal torque, and the torque from the
core--mantle interaction. The latter torque has four contributions,
$\langle {\bf T}_{viscous}\rangle,\,\langle {\bf
T}_{magnetic}\rangle$, $\langle{\bf T}_{topographic}\rangle$, and
$\langle{\bf T}_{pressure}\rangle$, for viscous, magnetic, topographic,
and pressure coupling, respectively.  The angled 
brackets indicate that these torques are averaged over an orbit
period; the core--mantle torques do not involve the orbital
elements, so they are intrinsically averaged.  We desire the variation of
${\bf L_m}$ relative to the precessing orbit system, where the
variation relative to inertial space is given by the total torque. We
therefore write $d{\bf L_m}/dt=\langle{\bf T_m}\rangle-\bs{\mu}\times {\bf
L_m}$, where $\bs{\mu}$ is the angular velocity of the orbit
precession.  If we
write ${\bf N_m}=\langle{\bf T_m}\rangle/C_m-\bs{\mu}\times\bs{\dot
\psi_m}$, we can equate each component of $(1/C_m)d{\bf L_m}/dt$ in
Eqs. (\ref{eq:dLdt}) to the corresponding component of ${\bf N_m}$ and
solve the resulting set for $d\dot\psi_m/dt,\;di_m/dt$, and
$d\Omega_m/dt$ to follow the motion of Mercury's mantle under
conservative and dissipative torques. We find
\begin{eqnarray}
\frac{d\dot\psi_m}{dt}&=&\sin{i_m}(N_{m{\s X}}\sin{\Omega_m}-N_{m{\s
Y}}\cos{\Omega_m})+N_{m{\s Z}}\cos{i_m},\nonumber\\
\frac{di_m}{dt}&=&-\frac{1}{\dot\psi_m}[\cos{i_m}(-N_{m{\s
X}}\sin{\Omega_m}+N_{m{\s Y}}\cos{\Omega_m})+N_{m{\s
Z}}\sin{i_m}],\nonumber\\ 
\frac{d\Omega_m}{dt}&=&\frac{1}{\dot\psi_m\sin{i_m}}(N_{m{\s
X}}\cos{\Omega_m}+N_{m{\s Y}}\sin{\Omega_m}).\label{eq:dpsidt}
\end{eqnarray}
We change variables to $p_m=\sin{i_m}\sin{\Omega_m}$ and
$q_m=\sin{i_m}\cos{\Omega_m}$ to eliminate the $\sin{i_m}$ singularity
in the third of Eqs. (\ref{eq:dpsidt}). Differentiating these variables
with respect to time, substituting the expressions for the time
derivatives from Eqs. (\ref{eq:dpsidt}), and expressing the circular
functions in terms of $p_m$ and $q_m$ yields
\begin{eqnarray}
\frac{d\dot\psi_m}{dt}&=&p_mN_{m{\s X}}-q_mN_{m{\s Y}}+
\sqrt{1-p_m^2-q_m^2}N_{m{\s Z}}\nonumber\\
\frac{dp_m}{dt}&=&\frac{1}{\dot\psi_m}\left[(1-p_m^2)N_{m{\s X}}
+p_mq_mN_{m{\s Y}}-p_m\sqrt{1-p_m^2-q_m^2}N_{m{\s Z}}\right]\nonumber\\
\frac{dq_m}{dt}&=&-\frac{1}{\dot\psi_m}\left[p_mq_mN_{m{\s
X}}+(1-q_m^2)N_{m{\s Y}}+q_m\sqrt{1-p_m^2-q_m^2}N_{m{\s Z}}\right]
\label{eq:finaleqs} 
\end{eqnarray}
as the equations to be solved for the behavior of the mantle under the
applied torques.  The corresponding equations for the core are
identical to Eqs. (\ref{eq:finaleqs}) with all the $m$ subscripts
replaced by $f$.  Now we must construct the expressions for
$\langle{\bf T}_{body}\rangle,\;\langle{\bf T}_{tide}\rangle,\,
\langle{\bf T}_{viscous}\rangle,\,\langle{\bf T}_{magnetic}\rangle$,
$\langle{\bf T}_{topographic}\rangle$, and $\langle{\bf
T}_{pressure}\rangle$  in terms of the variables of
Eqs. (\ref{eq:finaleqs}).
\subsection{Conservative gravitational torque}
The conservative body torque the Sun exerts on Mercury is
${\bf T}_{body}={\bf r}\times\nabla V$, where 
\begin{equation}
V=-\frac{Gmm_{\s\sun}}{r}\left[1-J_2\frac{R^2}{r^2}
\left(\frac{3\cos^2{\theta}}{2}-\frac{1}{2}\right) +
3C_{22}\frac{R^2}{r^2}\sin^2{\theta}\cos{2\phi}\right]
\label{eq:potential} 
\end{equation}
is the potential energy of the Sun in Mercury's gravitational field up to
second spherical harmonic degree, and $\bf r$ is the vector from
Mercury to the Sun.  In
Eq. (\ref{eq:potential}), $m$ and $m_{\s\sun}$ are the masses of
Mercury and the Sun, respectively, $G$ is the gravitational constant,
$R$ is Mercury's radius,
$\theta$ and $\phi$ are 
spherical polar coordinates relative to Mercury's principal axis
system of coordinates, which define the direction to the Sun, and
$J_2=[C-(A+B)/2]/mR^2$ and $C_{22}=(B-A)/4mR^2$ are unnormalized zonal
and tesseral coefficients of degree 2 in 
the spherical harmonic expansion of Mercury's gravitational potential.
$A<B<C$ are principal moments of inertia of Mercury.
It is expedient to form the cross product in the body system
of coordinates and express the resulting combinations of the spherical
polar coordinates in terms of scalar products to yield
\begin{eqnarray} 
({\bf r}\times\nabla V)_x&=&\frac{Gmm_{\s\sun}R^2}{r^5}({\bf r\cdot
k})({\bf r\cdot j})(3J_2-6C_{22}),\nonumber\\ 
({\bf r}\times\nabla V)_y&=&-\frac{Gmm_{\s\sun}R^2}{r^5}({\bf r\cdot
k})({\bf r\cdot i})(3J_2+6C_{22}),\nonumber\\ 
({\bf r}\times\nabla V)_z&=&12\frac{Gmm_{\s\sun}R^2}{r^5}({\bf r\cdot
i})({\bf r\cdot j})C_{22},\label{eq:torque1}
\end{eqnarray}
where ${\bf i,j,k}$ are unit vectors along the $x,y,z$ axes,
respectively. The 
components of ${\bf T}_{body}$ in the $XYZ$ orbit system are obtained
by rotations through the Euler angles $\Omega_m,i_m,\psi_m$
defined in Fig. \ref{fig:coordsystems}, where the scalar products in
Eqs. (\ref{eq:torque1}) are also expressed in terms of their $XYZ$
components. Then
\begin{eqnarray}
{\bf T}_{body}&=&-\frac{Gmm_{\s\sun}R^2}{r^3}{\Bigg\{}3J_2\sin{i_m}
\cos{i_m}\sin{(\omega+f-\Omega_m)}[\sin{(\omega+f)}{\bf I}-
\cos{(\omega+f)}{\bf J}]\nonumber\\
&&-1.5J_2\sin^2{i_m}\sin{2(\omega+f-
\Omega_m)}{\bf K}+\nonumber\\
&&6C_{22}\sin{i_m}[\cos{i_m}\cos{2\psi_m}\sin{(\omega+f- \Omega_m)}-\sin{2\psi_m}\cos{(\omega+f-\Omega_m)}]\times\nonumber\\
&&[-\sin{(\omega+f)}{\bf I}+\cos{(\omega+f)}{\bf J}]+\label{eq:torque2}\\
&&6C_{22}\left[-\left(\frac{3+\cos{2i_m}}{4}\right)\cos{2\psi_m}
\sin{2(\omega+f-\Omega_m)}+\cos{i_m}\sin{2\psi_m}\cos{2(\omega+f-\Omega_m)}\right]{\bf 
K}{\Bigg\}}, \nonumber
\end{eqnarray}
which is identical to Eq. (7) of Peale (2005) with $f$ replaced by
$\omega+f$ ($\omega=0$ in the 2005 paper). 

The average over the orbit is carried out by keeping all slowly varying
parameters constant, while the true anomaly and rotation are allowed
to vary.  With Mercury in the 3:2 spin orbit resonance,
$\dot\psi=1.5n+\dot\gamma$, where $\dot\gamma$ allows a small
variation in the rotation rate relative to the resonant value.
The stability of the resonance requires the  axis of
minimum moment of inertia (long axis) to be nearly aligned with the
direction to the Sun when Mercury is at perihelion. A slight deviation
from this condition will cause a free libration about this position
that will be damped by dissipation. We can represent the position of
the long axis by $\psi_m=1.5M+\omega-\Omega_m+\gamma$ from
inspection of Fig. \ref{fig:coordsystems}, where $\omega-\Omega_m$ is
the angular distance from the ascending node of the equator on the
orbit plane to the perihelion with $i_m\ll 1$, and $M$ is the mean
anomaly. The variations of $\omega$ and $\Omega$ are
sufficiently slow that they can be ignored on the orbital and rotation
time scales. At perihelion, $M=0$, so $\gamma$ is a small offset of the
long axis from the direction to the Sun when Mercury is at perihelion.
For the averaging procedure, $2\psi_m = 3M+2(\omega-\Omega_m)+
2\gamma$.  The rapidly varying quantities in Eq. (\ref{eq:torque2}) 
are thus $f,\,M$, and $r$ in terms which can  be isolated
by expanding the circular functions.  Non--zero averages of $a^3/r^3$,
$(a^3/r^3)\cos{2f}\cos{3M}$, $(a^3/r^3)\sin{2f}\sin{3M}$, and
$(a^3/r^3)\cos{3M}$, where $a$ is the semimajor axis of Mercury's
orbit, are expressed as series in the orbital eccentricity $e$, where
we truncate each series at $e^5$. After some algebraic manipulation
and combination of terms, we find the averaged torque to be
\begin{eqnarray}
\frac{\langle{\bf T}_{body}\rangle_{\s X}}
{C_m}&=&-\frac{n^2}
{\alpha_m}\Bigg[\frac{3}{2}J_2\sin{i_m}\cos{i_m}\cos{\Omega_m}
g_1(e)\nonumber\\ 
&&+\frac{3}{2}C_{22}\sin{i_m}(1+\cos{i_m})\cos{(\Omega_m
-2\gamma)}g_2(e)\nonumber\\
&&+\frac{3}{2}C_{22}\sin{i_m}(1+\cos{i_m})\cos{(2\omega-\Omega_m+
2\gamma)}g_3(e)\Bigg],\nonumber\\
\frac{\langle{\bf T}_{body}\rangle_{\s Y}}{C_m}&=&-\frac{n^2}
{\alpha_m}\Bigg[\frac{3}{2}J_2\sin{i_m}\cos{i_m}\sin{\Omega_m}
g_1(e)\nonumber\\ 
&&+\frac{3}{2}C_{22}\sin{i_m}(1+\cos{i_m})\sin{(\Omega_m
-2\gamma)}g_2(e)\nonumber\\
&&-\frac{3}{2}C_{22}\sin{i_m}(1+\cos{i_m})\sin{(2\omega-\Omega_m+
2\gamma)}g_3(e)\Bigg],\nonumber\\ 
\frac{\langle{\bf T}_{body}\rangle_{\s Z}}{C_m}&=&-\frac{n^2} 
{\alpha_m}\Bigg\{\frac{3}{2}C_{22}(1+\cos{i_m})^2
\Bigg[g_2(e)\sin{2\gamma}+g_4(e)\sin{(4\omega-4\Omega_m+
2\gamma)}\Bigg]\Bigg\},\label{eq:torque3}
\end{eqnarray}
where $C_m=\alpha_mmR^2$ defines $\alpha_m$, and $n^2=Gm_{\s\sun}/a^3$
has been used. The choice of $C_m$ in Eqs. (\ref{eq:torque3}) means
they apply to the mantle alone, and $J_2$ and $C_{22}$ correspond only
to this part of Mercury.  We show in Section 2.6 and Appendix B that
pressure torques effectively restore the full values of $J_2$ and
$C_{22}$ in these equations. In Eqs. (\ref{eq:torque3})
\begin{eqnarray}
g_1(e)&=&(1-e^2)^{-3/2}\nonumber\\
g_2(e)&=&\frac{7e}{2}-\frac{123e^3}{16}+\frac{489e^5}{128}+...\nonumber\\
g_3(e)&=&\frac{53e^3}{16}+\frac{393e^5}{256}+...\nonumber\\
g_4(e)&=&\frac{85e^5}{2560}+...\label{eq:ge}
\end{eqnarray}
There are additional terms in the $X$ and $Y$
components of ${\bf T}_{body}$ to order $e^5$ with factors
$\sin{i_m}(1-\cos{i_m})$. Since Mercury's obliquity is about 2 arcmin
(Margot {\it et al.}, 2007, 2012) and since we always start the
integrations close to the final state, these factors coupled with
lowest-order factors of $e^3$ or $e^5$ make these terms negligibly
small compared with the terms that are retained. With core parameters,
Eqs. (\ref{eq:torque3}) apply to the core as well, but we show below
that pressure forces between core and mantle effectively cancel the
gravitational torque on the core.
\subsection{Tidal torque}
We choose the simplest tidal model, in which an equilibrium tidal
distortion has its maximum displacement at a point that was directly
under the disturbing body, here the Sun, a short time $\Delta t$ in
the past. This model is equivalent to setting the tidal dissipation
factor $Q$ inversely proportional to frequency. This model is not a good
representation of the behavior of solid materials  
(Castillo-Rogez {\it et al.}, 2011), but since the frequencies involved
will be near the orbital frequency $n$ and are confined to a fairly narrow
range, we expect that the time scale for evolution to equilibrium will
depend more on the value of $Q$ than on the tidal model. We are
interested in the final equilibrium state and not necessarily in the
rate of approach.
It can be shown that for a tidal frequency equal
to the orbital mean motion $n$, $\Delta t=1/(Qn)$,  where $Q$ has the
value appropriate to a tidal frequency $n$ ({\it e.g.,} Peale 2005,
2006). There are many treatments in which this model is developed and
used (e.g., Mignard 1979, 1980, 1981; Hut 1981; Peale 2005, 2007).  
Peale (2005) derived the tidal torque averaged over an orbit period
for the case where the  argument of periapse $\omega$ is set to zero.
The averaged 
equations must be re-derived by procedures in the 2005 paper to include
non--zero values of $\omega$.  We find
\begin{eqnarray}
\frac{\langle{\bf T}_{tide}\rangle}{C_m}&=&3\frac{n^2}{\alpha_m}
\frac{m_{\s\sun}}{m}\frac{R^3}{a^3}k_2\Delta t
\Bigg\{-\dot\psi_m\bigg[
p_m\frac{f_2(e)}{2}+(q_m\sin{2\omega}-p_m\cos{2\omega})\frac{f_3(e)}{2}\bigg]{\bf
I}\nonumber\\
&&+\dot\psi_m\bigg[q_m\frac{f_2(e)}{2}+(q_m\cos{2\omega}+p_m\sin{2\omega})
\frac{f_3(e)}{2}\bigg]{\bf J}\nonumber\\
&&+\left[nf_1(e)-f_2(e)\dot\psi_m\sqrt{1-p_m^2-q_m^2}\right]{\bf K}\Bigg\}
\label{eq:tidetorque} 
\end{eqnarray}
where $k_2$ is the second-degree potential Love number, and where we
have converted the circular functions in the expressions to 
our variables $p_m$ and $q_m$. In Eqs. (\ref{eq:tidetorque})
\begin{eqnarray}
f_1(e)&=&\left(1+\frac{15e^2}{2}+\frac{45e^4}{8}+
\frac{5e^6}{16}\right){\Bigg/}(1-e^2)^6\nonumber\\ 
f_2(e)&=&\left(1+3e^2+\frac{3e^4}{8}\right){\Bigg/}
(1-e^2)^{9/2}\nonumber\\
f_3(e)&=&\left(\frac{3e^2}{2}+\frac{e^4}{4}\right){\Bigg/}
(1-e^2)^{9/2}\label{eq:fe}
\end{eqnarray}
\subsection{Viscous core--mantle torque}
For the viscous interaction between the core and mantle we assume that
the torque is simply proportional to the difference in the vector angular
velocities.  This assumption is consistent with the core rotating as a
rigid body (Poincar\'e, 1910).
\begin{eqnarray}
\langle{\bf T}^m_{viscous}\rangle&=&-\beta(\bs{\dot\psi}_m-
\bs{\dot\psi}_f),\nonumber\\
\langle{\bf T}^f_{viscous}\rangle&=&-\langle{\bf T}^m_{viscous}\rangle
\label{eq:coretorq1}
\end{eqnarray}
are the torques on the mantle and core, respectively. Since the
core--mantle torques do not change with orbital position, the averaged
values are the same as the defining expressions.
If we assume that the angular velocities are parallel and that the torques
in Eqs. (\ref{eq:coretorq1}) are the only ones acting, the time constant
for an exponential decay of the differential rotation is
$C_fC_m/[\beta(C_f+C_m)]$, which follows from the difference in the
angular momentum equations for the mantle and core appropriate to the
above torques. We connect $\beta$ to the viscosity of the
fluid by equating this time constant to that for the decay of fluid rotation
relative to its container.  We thereby restore the effects of
the fluid nature of the core. If the viscosity is small, a time scale
for damping the rotation of the 
core relative to the CMB is $R_f/(\dot\psi \nu)^{1/2}$ (Greenspan and
Howard, 1963), where $\nu$ is the kinematic viscosity, and $R_f$ is the
radius of the fluid core. The
time scale is derived for no density stratification in the core and
depends on circulation of the fluid through the Ekman boundary layer, the
thickness of which is small compared with the core radius.  If the
viscosity is large, the latter condition is not necessarily satisfied,
and the viscous time scale $R_f^2/\nu$ becomes appropriate.  It is
probably unlikely that the viscosity of the liquid core material is uniform
throughout or that there is no density stratification therein.  A
purely viscous time scale using the viscosity at the CMB may therefore
be more appropriate than the Greenspan and Howard time scale, but we
determine the dependence of the coupling constant $\beta$ on the
liquid core viscosity for both possibilities.

\begin{equation}
\beta=\frac{(\dot\psi\nu)^{1/2}}{R_f}\frac{C_fC_m}{C_f + C_m}
\label{eq:beta1}
\end{equation}
or
\begin{equation}
\beta=\frac{\nu}{R_f^2}\frac{C_fC_m}{C_f+C_m}
\label{eq:beta2}
\end{equation}
for small and large viscosities, respectively. 
With $\bs{\dot\psi}_m=\dot\psi_m{\bf k}_m=\dot\psi_m[p_m{\bf I}-q_m{\bf
J}+\sqrt{1-p_m^2-q_m^2}{\bf K}]$ and with a similar expression for
$\bs{\dot\psi}_f$, we can write
\begin{eqnarray}
\langle T^m_{viscous}\rangle_{\s X}&=&-\beta(p_m\dot\psi_m-
p_f\dot\psi_f),\nonumber\\
\langle T^m_{viscous}\rangle_{\s Y}&=&-\beta(-q_m\dot\psi_m+
q_f\dot\psi_f),\nonumber\\
\langle T^m_{viscous}\rangle_{\s Z}&=&-\beta\left(\dot\psi_m
\sqrt{1-p_m^2-q_m^2}-\dot\psi_f \sqrt{1-p_f^2-q_f^2}\right), 
\label{eq:tmcomponents}
\end{eqnarray}
where the same equations give the components of $\langle{\bf
T}^f_{viscous}\rangle$ but with the leading signs reversed. 

\subsection{Magnetic core--mantle torque}

An expression for the magnetic torque on the mantle is given by
Buffett (1992).
\begin{equation}
{\bf T}_{magnetic}=\frac{|\Phi_{\s B}|}{\mu_0}\int_{\s S}[B_r(r)]^2
[r^2{\bf w}-({\bf r}\cdot{\bf w}){\bf r} ]dS, \label{eq:tmag}
\end{equation}
where the integral is over the CMB and where ${\bf
w}=\bs{\dot\psi}_m-\bs{\dot\psi}_f$ is the relative 
angular velocity between the mantle and core, $B_r(r)$ is the radial
component of the magnetic field at the CMB, $|\Phi_{\s B}|\approx 43$ s/m
when the thickness of the conducting layer in the mantle exceeds 200
m with conductivity $\sigma=5\times 10^5$ S/m appropriate to Buffett's
assumed properties of the Earth's mantle at the CMB, and
$\mu_0=4\pi\times 10^{-7}\,{\rm N/A^2}$ is the permeability of
free space. The magnetic torque will of course change with different
values of the thickness of the conducting layer and electrical
conductivity, but the changes in $|\Phi_{\s B}|$ are easily determined
within the Buffett (1992) theory.

The magnetic field of Mercury is primarily a spin--aligned dipole,
for which the radial component is given by
\begin{equation}
B_r(r,\theta)=\frac{\mu_0\mathcal{M}\cos{\theta}}{2\pi r^3},\label{eq:Br}
\end{equation}
where $\theta$ is the co-latitude and $\mathcal{M}$ is the magnetic
dipole moment. The dipole is here centered at the
origin of the coordinate system. But the center of the Mercury's dipole
is offset northward from the center of the planet  by
486 km (Anderson {\it et al.} 2011, 2012), and this offset complicates the
calculation of $B_r$ relative to 
the center of Mercury. We calculate the magnetic
torque evolution of the system for a Mercury-centered dipole, with
straightforward  integration over the CMB. With ${\bf
r}=R_f{\hat{\bf r}}$, where $\hat{\bf r}$ is a unit vector in the
direction of $\bf r$, substitution of Eq. (\ref{eq:Br}) into
Eq. (\ref{eq:tmag}) yields
\begin{equation}
{\bf T}_{magnetic}=\frac{|\Phi_{\s B}|\mu_0\mathcal{M}^2}{4\pi^2R_f^2}
\int_0^{2\pi}\int_0^{\pi}[{\bf w}-(\hat{\bf r}\cdot {\bf w}){\hat{\bf
r}}] \cos^2{\theta}\sin{\theta}d\theta d\phi, \label{eq:inttmag}
\end{equation}
where $\theta$ and $\phi$ are spherical polar coordinates of a point
on the CMB in the body xyz
system. The relative angular velocity $\bf w$ remains constant for the
integration over the CMB.  We find
\begin{equation}
{\bf T}_{magnetic}=\frac{|\Phi_{\s B}|\mu_0\mathcal{M}^2}{15\pi
R_f^2}[4(w_x{\bf i}+w_y{\bf j}+w_z{\bf k})-2w_z{\bf k}],\label{eq:tmag2}
\end{equation}
where ${\bf i},{\bf j},{\bf k}$ are unit vectors along the xyz body
axes, respectively.
The vector $\bf w$ in Eq. (\ref{eq:tmag2}) is just
$\bs{\dot\psi}_m-\bs{\dot\psi}_f$ which can be expressed in the XYZ
orbit system, and the last term is just $2({\bf w}\cdot {\bf k}_m)
{\bf k}_m$, which is also expressible in the XYZ system. (Recall that ${\bf
k}_m=p_m{\bf I}-q_m{\bf J}+\sqrt{1-p_m^2-q_m^2}{\bf K}$.)
There results

\begin{eqnarray*}
T_{magnetic}^{\s X}&=&\frac{|\Phi_{\s B}|\mu_0\mathcal{M}^2}{15\pi R_f^2}
\bigg\{4(p_m\dot\psi_m-p_f\dot\psi_f)-2p_m\bigg[(p_m\dot\psi_m-p_f\dot\psi_f)p_m
+ (q_m\dot\psi_m-q_f\dot\psi_f)q_m+\\
&&\left(\sqrt{1-p_m^2-q_m^2}\dot\psi_m 
-\sqrt{1-p_f^2-q_f^2}\dot\psi_f\right)\sqrt{1-p_m^2-q_m^2}\bigg]\bigg\}
\end{eqnarray*}
\begin{eqnarray*}
T_{magnetic}^{\s Y}&=&\frac{|\Phi_{\s B}|\mu_0\mathcal{M}^2}{15\pi R_f^2} 
\bigg\{-4(q_m\dot\psi_m-q_f\dot\psi_f)+2q_m\bigg[(p_m\dot\psi_m-p_f\dot\psi_f)p_m
+ (q_m\dot\psi_m-q_f\dot\psi_f)q_m+\nonumber\\
&&\left(\sqrt{1-p_m^2-q_m^2}\dot\psi_m
-\sqrt{1-p_f^2-q_f^2}\dot\psi_f\right)\sqrt{1-p_m^2-q_m^2}\bigg]\bigg\}\nonumber
\end{eqnarray*}
\begin{eqnarray}
T_{magnetic}^{\s Z}&=&\frac{|\Phi_{\s B}|\mu_0\mathcal{M}^2}{15\pi R_f^2} 
\bigg\{4(\sqrt{1-p_m^2-q_m^2}\dot\psi_m-\sqrt{1-p_f^2-q_f^2}\dot\psi_f)- 
\nonumber\\
&&2\sqrt{1-p_m^2- q_m^2}\bigg[(p_m\dot\psi_m-p_f\dot\psi_f)p_m
+ (q_m\dot\psi_m-q_f\dot\psi_f)q_m+\nonumber\\
&&\left(\sqrt{1-p_m^2-q_m^2}\dot\psi_m
-\sqrt{1-p_f^2-q_f^2}\dot\psi_f\right)\sqrt{1-p_m^2-q_m^2}\bigg]\bigg\}
\label{eq:tmag3}
\end{eqnarray}  

\subsection{Topographic core--mantle torque}

The differential rotation of the mantle and core can lead to dynamic 
pressure forces on any topography on the CMB. 
The form of the torque on topography starts with the dynamic pressure of a
fluid times the area element $\rho u^2dS$, where $u$ is the
magnitude of a relative velocity, and $\rho$ is a fluid density. 
This gives a force on dS as if   
the relative fluid velocity were impinging on the CMB in the normal
direction. If the vertical extent of the bump that occupies area
element dS on the CMB leads 
to a cross section as seen by the fluid that is flowing parallel 
to the mean CMB surface of $dS\sin{\delta}$, where
angle $\delta$ is the slope of the bump relative to the mean CMB, 
the dynamic force on the bump is $\rho
u^2\sin{\delta}dS$. But this element of surface results not from a face
perpendicular to the flow but from the projection of the slanted
surface perpendicular to the direction of $\bf u$.  So a factor
$\zeta<1$ is inserted as an unknown efficiency of the transfer of
linear momentum to the slanted surface. The efficiency factor $\zeta$
is difficult to estimate, so we leave it as a parameter. 

If the velocity of the CMB relative to the fluid is represented by
${\bf u}=(\bs{\dot\psi}_m-\bs{\dot\psi}_f)\times {\bf R}_f$, and
$\sin{\delta}=|\hat{\bf r}\times \hat{\bf n}|$, with $\hat{\bf n}$
a unit vector normal to the CMB surface at ($R_f,\theta,\phi$) and
$\hat {\bf r}$ a unit vector in the direction of $\bf r$, then the
increment of topographic pressure force on the mantle is 
\begin{equation} 
-\rho_f u\zeta\sin{\delta}dS{\bf u}, \label{eq:stress}
\end{equation}
where $\rho_f$ is the density of the liquid core at the CMB. 
If the surface is spherical, the stress is zero. (We do not consider
the effect of the ellipsoidal mean surface here.) 
The force $d{\bf F}$ on the mantle from the stress on an area $dS$ is
in the direction of $-\bf u$, and the torque on the mantle from $d{\bf
F}$ is ${\bf R}_f\times d{\bf F}$. 

The total torque on the mantle from topography on the CMB is then
\begin{equation}
{\bf T}_{topographic}= -\int_0^{2\pi}\int_0^\pi \zeta \sin{\delta}
\rho_fu{\bf R}_f\times {\bf u}R_f^2\sin{\theta}d\theta d\phi.
\label{eq:ttop} 
\end{equation}
In the integrand of Eq. (\ref{eq:ttop}), $u=\sqrt{{\bf u}\cdot{\bf u}}$
prohibits the analytic integration over the CMB. However, with
$\bs{\dot\psi}_m=\dot\psi_m(p_m{\bf I}-q_m{\bf J}+
\sqrt{1-p_m^2-q_m^2}{\bf K})$ and a similar expression for
$\bs{\dot\psi}_f$, and with the components of ${\bf R}_f$
in the same system of coordinates expressed in spherical polar
coordinates, it is easy to express $u^2$ in terms of the $p$s and $q$s
and $\theta$ and $\phi$. If we limit ourselves to small obliquities
in the final evolution of the system to equilibrium, then $p_m,q_m,p_f,q_f\ll
1$. With these quantities set to zero in the expression for $u^2$, we
have $u\approx R_f|\bs{\dot\psi}_m-\bs{\dot\psi}_f|\sin{\theta}$ and
Eq. (\ref{eq:ttop}) is integrable to yield
\begin{eqnarray}
\frac{T_{topographic}^{\s X}}{C_m}&=&
-\chi|\bs{\dot\psi}_m-\bs{\dot\psi}_f|(p_m\dot\psi_m-p_f\dot\psi_f),
\nonumber\\
\frac{T_{topographic}^{\s
Y}}{C_m}&=&\chi|\bs{\dot\psi}_m-\bs{\dot\psi}_f |
(q_m\dot\psi_m-q_f\dot\psi_f),\nonumber \\
\frac{T_{topographic}^{\s Z}}{C_m}&=&-2\chi|\bs{\dot\psi}_m-
\bs{\dot\psi}_f|
(\sqrt{1-p_m^2-q_m^2}\dot\psi_m-\sqrt{1-p_f^2-q_f^2}\dot\psi_f),
\label{eq:ttop2} 
\end{eqnarray}
where $\chi=5\zeta\sin{\delta}\pi^2\rho_fR_f^5/(16\alpha_mmR^2)=(15/64) 
(m_f/m)(1/\alpha_m)(R_f^2/R^2)\pi\zeta\sin{\delta}=
2.4301\zeta\sin{\delta}$, with $m_f/m=0.7334$, $R_f=1998\,{\rm
km}$, and $\alpha_m=0.149$ for the two-layer model discussed in the
next section. We have assumed
that the ``roughness'' of the CMB is the same everywhere such that
$\zeta \sin{\delta}$ is a constant.

\subsection{Pressure torque}
The influence of pressure torques at the CMB on Mercury's libration in
longitude have been considered by Van Hoolst {\it et al.} (2012).
Here we generalize the pressure torques to three dimensions, where
variations in fluid velocity are important. Until now we have ignored
the presence of the solid inner core, and we will ultimately determine
the pressure torque for an entirely fluid core.  However, in
determining the shape of the CMB as an equipotential surface, it is
convenient to also determine the shape of the inner core boundary
(ICB) as a function of the inner core size and density. We will use
the shape of 
the ICB in a later work determining the equilibrium spin of the inner
core and its gravitational effect on the mantle spin. We therefore
model the interior structure of Mercury as three homogeneous layers,
mantle plus crust, fluid outer core, and solid inner core. The observed
values of $J_2$ and $C_{22}$ can be expressed as a sum of
contributions from the ellipsoidal shapes of the surfaces of each
layer. For principal axes of the ellipsoids given by $a>b>c$ and
principal moments of inertia $A<B<C$, $J_2$ is expressed in terms of
the mean polar ellipticities $\epsilon=(\epsilon_a+\epsilon_b)/2=[(a-c)/r_0+
(b-c)/r_0]/2$ ($r_0=$ mean radius), the surface radii
$R_m=R,\,R_f,\,{\rm and}\,R_s$ ($R_s$ is the solid 
inner core radius), and the densities of
the layers, $\rho_m,\,\rho_f, \,{\rm and}\,\rho_s$. $C_{22}$ is
expressed in terms 
of the equatorial ellipticities $\xi=(a-b)/r_0$, the densities, and
the surface radii. The subscripts $m,\,f,\,{\rm and}\,s$ refer to the
mantle, fluid outer 
core, and solid inner core, respectively.  Additional equations  
in these same variables are determined by the assumption of
hydrostatic equilibrium, where the CMB and the
inner core boundary 
(ICB) are equipotential surfaces. In computing the gravitational
potential throughout the interior of the planet, we assume that the
surfaces have the form
\begin{equation}
r=r_0\left[1-\frac{2\epsilon}{3}P_{20}(\cos{\theta})+
\frac{\xi}{6}P_{22}(\cos{\theta})\cos{2\phi}\right],\label{eq:rsurface}
\end{equation}
where $P_{ij}$ are Legendre functions, and $\theta$ and $\phi$ are spherical
polar coordinates ($\theta=$ colatitude) relative to the 
principal axis system in which the $x$ axis is the axis of minimum
moment of inertia (semiaxis $a$). The coefficients of the Legendre
functions in Eq. (\ref{eq:rsurface}) are determined
by evaluating coefficients $\Delta R_1/r_0$ and $\Delta
R_2/r_0$ of the two Legendre functions by setting $r=a,\,b,\,{\rm and}\,c$ for
appropriate choices of the angles therein and solving for the $\Delta
R_i$ in terms of differences in the axis lengths expressed by $\epsilon$ and
$\xi$.  

The equipotential surfaces at the CMB and ICB give two
equations in $\epsilon_i$ and $\xi_i$, which can be decomposed into
equations for the $\epsilon_i$ and separately for the $\xi_i$
because of the orthogonality of the Legendre functions. $J_2$ and 
the equations of two equipotential surfaces involving the polar
ellipticities $\epsilon_i$ comprise three equations with additional
parameters, $\rho_i$ and $R_i$. Similarly, three equations are found
for $C_{22}$ and two equipotential surfaces involving
$\xi_i,\,\rho_i,\,{\rm and}\,R_i$. The densities $\rho_i$ and 
the radii $R_i$ are constrained by observed values of $C/mR^2$,
$C_m/C$, and total mass $m$. These are three equations in the six
unknowns $\rho_i$ and $R_i$. But $R_m=R$, and if we specify the inner
core parameters $\rho_s$ and $R_s$, the three remaining variables
$\rho_f$, $\rho_m,$ and $R_f$ are determined.
If we use the values of these parameters
so derived along with the assumptions for $\rho_s\,{\rm and}\,R_s$ in the
$\epsilon$ and $\xi$ equations, the latter can be solved uniquely for
the $\epsilon_i$ and $\xi_i$.  The internal structure so obtained is
consistent with the observables $J_2=5.03\times
10^{-5},\,C_{22}=0.809\times 10^{-5},\,C/mR^2=0.346,\,C_m/C=0.431,\, {\rm
and}\,M=3.301\times 10^{26}$g. The values of $J_2,\,C_{22},$ and $M$
come from Smith {\it et al.} (2012), and $C/mR^2$ and $C_m/C$ from
Margot {\it et al.} (2012). In summary there are 12 unknowns 
($3\epsilon$s, $3\xi$s, $3\rho$s, $3R$s) and 10 equations
($C/mR^2,\,C_m/C,\,m,\,R,\,J_2,\,C_{22}$, two equipotential surface
conditions at the CMB and two at the ICB). We can specify the inner
core radius $R_s$ and density $\rho_s$ reducing the unknowns to
10. This exercise is carried out in Appendix A,
and results  are given in Table \ref{tab:alphabeta} for an inner
core of radius $0.6R$ and density $8\,{\rm g/cm^3}$ and
alternatively, for no inner core. The ICB may be more likely to be an
equipotential surface than the CMB.  In evaluating the consequences of
the pressure torque on the equilibrium spin axis positions for an
axially symmetric CMB with no solid inner core, we consider both the
equipotential value of $\epsilon_f=7.161\times 10^{-5}$ given in Table
\ref{tab:alphabeta} and double this value, where the latter is a crude
measure of the effects of the CMB not being an equipotential surface.
\begin{table}[h]
\caption{Interior densities, outer core radii, and ellipticities for two
choices of solid inner core properties. The measured ellipticities
from MESSENGER spacecraft data are provided by M. E. Perry and
R.J. Phillips (private communication, 2013), and the rotational ellipticity
is that of the rotational equipotential surface at the CMB.}
\label{tab:alphabeta}
\centering 
\begin{tabular}{|c|c|c|c|c|}
\multicolumn{5}{c}{Densities}\\
\hline
$R_s/R$&$\rho_s\,{\rm g/cm^3}$&$\rho_{\s f}\,{\rm g/cm^3}$&$\rho_m\,{\rm
g/cm^3}$&$R_f\,{\rm km}$\\ 
\hline
0.6&8.0&6.510&3.347&2027\\
\hline
0.0&-&7.254&3.203&1998\\
\hline
\end{tabular}
\vspace{.4in}
\begin{tabular}{|c|c|c|c|c|c|c|}
\multicolumn{7}{c}{Ellipticities}\\
\hline
$R_s/R$&$\epsilon_s$&$\epsilon_f$&$\epsilon_m$&$\xi_s$&$\xi_f$&$\xi_m$\\
\hline
0.6&$6.818\times 10^{-5}$&$6.952\times 10^{-5}$&$1.755\times
10^{-4}$&$4.214\times 10^{-5}$&$3.396\times 10^{-5}$&$1.170\times 10^{-4}$\\
\hline
0.0&-&$7.161\times 10^{-5}$&$1.797\times 10^{-4}$&-&$4.607\times
10^{-5}$&$1.156\times 10^{-4}$\\
\hline
\multicolumn{2}{|c|}{Measured}&-&$\sim 6.86\times 10^{-4}$&-&-&$\sim
5.07\times 10^{-4}$\\
\hline
\multicolumn{2}{|c|}{Rotation}&$3.79\times 10^{-7}$&-&-&-&-\\
\hline
\end{tabular}
\end{table}
Also shown in Table \ref{tab:alphabeta} are the measured surface
ellipticities  and the rotational ellipticity of the CMB.  Because of
Mercury's slow rotation, the centrifugal potential
(Eq. (\ref{eq:centrif})) has a negligible effect on the overall values
of $\epsilon_i$ and is neglected in the solution for the
ellipticities. Similarly, the averaged solar potential
(Eq. (\ref{eq:sunpotential})), comparable in magnitude to the
centrifugal potential, is also neglected. The large discrepancies between 
the observed surface ellipticities and those consistent with $J_2$ and
$C_{22}$ might be removed by accounting for the distribution of
lower-density crustal material. 

The torque on the mantle due to the fluid pressure $P$ at the CMB is 
\begin{equation}
{\bf T}_{\s P}=\iint_S{\bf r}\times{\bf n}PdS=\iiint_V{\bf
r\times \nabla} P dV, \label{eq:presstorque}
\end{equation}
where $\bf r$ is the vector from the center of Mercury to a point on
the CMB, and $\bf n$ is the normal to the CMB directed outward at that
point. The surface integral is over the CMB, and the volume integral,
which follows from the divergence theorem, is over the fluid core. The
pressure torque is zero for a spherically symmetric CMB, but a
hydrostatic CMB is distorted by the asymmetric gravitational field and
only slightly by the centrifugal and solar potentials.    

For simplicity we evaluate the pressure torque under the assumption
that Mercury's CMB is axially symmetric.  This assumption allows
procedures in the literature for the Earth's core--mantle boundary to
be used (e.g., Melchior, 1986), and we will see
that the pressure torque is so dominant that refinements from axial
asymmetry cannot change the overall results but seriously
complicate the analysis. The usefulness of the volume integral
expression for the pressure torque comes from the relations
\begin{eqnarray}
\nabla P&=&-\rho_f\nabla\Phi-\rho_f\frac{d{\bf v}}{dt},\nonumber\\
\frac{d{\bf v}}{dt}&=&\frac{\partial{\bf v}}{\partial t}+({\bf
v}\cdot\nabla){\bf v},\label{eq:delp}
\end{eqnarray}
where $\Phi$ is the gravitational potential and $\bf v$ is the
velocity of the core fluid, which satisfies $\bf v\cdot n=0$. The first
equation is a form of Newton's second law in which pressure and
gravity forces are 
expressed per unit volume ($-\nabla P-\rho_f\nabla\Phi$). The second
equation invokes the material derivative and accounts for
advection. We assume steady precession in the orbit frame so that
$\partial{\bf v}/\partial t=0$. (Note that Eqs. (\ref{eq:delp}) are
written under the assumption that the $x'y'z'$ system is inertial,
when in fact the orbit frame is precessing.  However, since the time
scale for the orbital precession is so much longer than that of the
spin precession within the orbit system, the deviations from $\partial
{\bf v}/\partial t$ in the orbit frame from 0 are negligibly small.)
The potential term in Eq.(\ref{eq:delp}) consists of the external
potential due to the Sun and the internal potentials from the
distorted layers.  However, the latter can exert a gravitational
torque only if the layers are misaligned. We ignore the pressure
torques from the internal potentials.  The pressure torque from the
external potential is evaluated in Appendix B, where it is shown to
add to the gravitational torque on the mantle as if a thin layer of
density $\rho_f$ that lies outside the largest sphere that fits inside
the CMB were included with the mantle in calculating the gravitational
torque on the latter.  That fluid layer is precisely the core
contribution to $J_2$ and $C_{22}$, so if there is no inner core or it
is otherwise neglected, the torque on the mantle due to the Sun is for
the full values of $J_2$ and $C_{22}$ to account for that contribution
to the pressure torque.  Only the $({\bf
v}\cdot\nabla){\bf v}$ term in $\nabla P$ remains to be determined.
 
We modify the procedure outlined by Melchior (1986) and 
approximate the CMB surface
given by Eq. (\ref{eq:rsurface}) with $\xi=0$ as the ellipsoid with
semiaxes $a>c$.  
\begin{equation}   
x'^2+y'^2+\frac{a^2}{c^2}z'^2=x'^2+y'^2+f_{\epsilon}z'^2=a^2,\label{eq:ellipsoid}
\end{equation}
where $x',y',z'$ are non--rotating coordinates, $x'y'$ is the equator
plane of Mercury, and $f_\epsilon=1+2\epsilon$ to first order in $\epsilon$,
where $\epsilon=(a-c)/r_0$ is the ellipticity that appears in
Eq. (\ref{eq:rsurface}). The primes on $x',y',z'$ distinguish these
coordinates from the rotating coordinates $x,y,z$. (Note that this
equation could not be written in the non rotating frame for an axially
asymmetric CMB, which is the motivation for choosing the axially
symmetric case.) The normal to this 
surface is given by  
\begin{equation}
{\bf n'}=\frac{\nabla a^2}{|\nabla a^2|}=\frac{x'{\bf i'}+y'{\bf
j'}+f_\epsilon z'{\bf k'}}{\sqrt{x'^2+y'^2+f_\epsilon^2 z'^2}}
\label{eq:normalellisoid} 
\end{equation}
We can transform the ellipsoid to a homologous Poincar\'e sphere with
$x''=x',\,y''=y',\,z''=f_\epsilon z'$, where ${\bf n''}=(x''{\bf i'}
+y''{\bf j'}+z''{\bf
k'})/r_0$ is the normal to the sphere.  Consistent with the Poincar\'e
(1910) result that the fluid core has a uniform vorticity (rigid body
rotation), the fluid velocity in the sphere is
\begin{equation}
{\bf v''}=\bs{\dot\psi}_f\times{\bf
r}=(\dot\psi_{fy'}z''-\dot\psi_{fz'}y''){\bf i'}
+(\dot\psi_{fz'}x''-\dot\psi_{fx'}z''){\bf j'}
+(\dot\psi_{fx'}y''-\dot\psi_{fy'}x''){\bf k'},
\end{equation}
which becomes, after converting back to the ellipsoid,
\begin{equation} 
{\bf v}=(\dot\psi_{fy'}z'-\dot\psi_{fz'}y'){\bf i'}
+(\dot\psi_{fz'}x'-\dot\psi_{fx'}z'){\bf j'}
+\frac{(\dot\psi_{fx'}y'-\dot\psi_{fy'}x')}{f_\epsilon}{\bf k'},\label{eq:v}
\end{equation}
which is the velocity in the ellipsoid. Note that the condition ${\bf v\cdot
n'}=0$ is satisfied and that $\nabla\cdot{\bf v}=0$.
\begin{equation}
({\bf v\cdot\nabla}){\bf v}=(v_{z'}\dot\psi_{fy'}-v_{y'}\dot\psi_{fz'}){\bf i'}
+(v_{x'}\dot\psi_{fz'}-v_{z'}\dot\psi_{fx'}){\bf j'}+
\frac{(v_{y'}\dot\psi_{fx'}-v_{x'}\dot\psi_{fy'})}{f_\epsilon}
{\bf k'}.\label{eq:vdotdelv} 
\end{equation} 
With the $\nabla\Phi$ contribution to $\nabla P$ accounted for
earlier, we can write $\nabla 
P=-\rho_f({\bf v\cdot\nabla}){\bf v}$ and integrate each 
component of the gradient as a guide to $P(x',y',z')$. The function
\begin{equation}
P=-\rho_f\left[\frac{x'y'\dot\psi_{fy'}\dot\psi_{fx'}}{f_\epsilon}+
z'x'\dot\psi_{fz'}\dot\psi_{fx'}+z'y'\dot\psi_{fz'}\dot\psi_{fy'}-
(x'^2+z'^2)\frac{\dot\psi_{fy'}^2}{2f_\epsilon}- 
(x'^2+y'^2)\frac{\dot\psi_{fz'}^2}{2}-
(y'^2+z'^2)\frac{\dot\psi_{fx'}^2}{2f_\epsilon}
\right]\label{eq:p}
\end{equation}
produces the components of $\nabla P$ except for $f_\epsilon$ missing in the
denominator of two terms in the $\bf k$ component of the
gradient. To first order in $\epsilon$ with $r_0\rightarrow R_f$, 
\begin{equation}
{\bf r}\times{\bf n'}=\frac{f_\epsilon-1}{R_f}(y'z'{\bf i'}-x'z'{\bf
j'})\label{eq:rxn1}
\end{equation}
contains the factor $\epsilon$, so $f_\epsilon$ in the denominator becomes
1 in the integrand of Eq. (\ref{eq:presstorque}) to first order in
$\epsilon$.  Integration of Eq. (\ref{eq:presstorque}) after
substitution of Eqs.(\ref{eq:p}) and (\ref{eq:rxn1}) yields
\begin{equation}
{\bf T}_{\s P}=-\frac{8\pi}{15}\rho_f\epsilon_f R_f^5\dot\psi_{fz'}
(\dot\psi_{fy'}{\bf i'}-\dot\psi_{fx'}{\bf j'}),\label{eq:presstorque2}
\end{equation} 
where the contribution to ${\bf T}_{\s P}$ from the external potential
is not included here, but the contribution by the latter follows from
the assumption that the gravitational torque on the mantle plus crust
e corresponds to the full values of $J_2$ and $C_{22}$ (Appendix B). 

The pressure torque on the core is the negative of that on the
mantle.  The potential part of the pressure torque on the mantle is
equivalent to the gravitational torque on the thin layer of core fluid
outside the largest sphere that fits inside the CMB. The negative of
this torque,  part of the pressure torque on the core, exactly cancels
the direct gravitational torque on the core, and we need not include
either this part of the pressure torque on the core or the direct
gravitational torque on the core in the coupled equations. The
negative of Eq. (\ref{eq:presstorque2}) is thereby the only torque on
the core.   
Eq. (\ref{eq:presstorque2}) must be written in terms of the variables
of Eq. (\ref{eq:finaleqs}). Recall that $\bf i'$ and $\bf j'$ are unit
vectors in Mercury's equator plane that do not rotate with the
planet. The term in the parentheses in Eq. (\ref{eq:presstorque2}) is
$-{\bf k'}\times \bs{\dot\psi}_f$, so the choice of the $x'y'$ axes is
arbitrary. We choose the $x'$ axis to lie along the node of the equator
on the orbit plane, such that
\begin{eqnarray}
{\bf i'}&=&\frac{q_m}{\sqrt{q_m^2+p_m^2}}{\bf I}+
\frac{p_m}{\sqrt{p_m^2+q_m^2}}{\bf J}\nonumber\\
{\bf j'}&=&-\frac{p_m\sqrt{1-p_m^2-q_m^2}}{\sqrt{p_m^2+q_m^2}}{\bf I}+
\frac{q_m\sqrt{1-q_m^2-p_m^2}}{\sqrt{p_m^2+q_m^2}}{\bf J}+
\sqrt{p_m^2+q_m^2}{\bf K}, \label{eq:bfibfj}
\end{eqnarray} 
where definitions of $p_m$ and $q_m$ in terms of $i_m$ and $\Omega_m$
are used.  With ${\bf k'}=p_m{\bf I}-q_m{\bf
J}+\sqrt{1-p_m^2-q_m^2}{\bf K}$ and a similar expression for ${\bf
k'_f}$ and with $\dot\psi_{fz'}=\dot\psi_f{\bf k'}_f\cdot{\bf
k'},\,\dot\psi_{fx'}=\dot\psi_f{\bf k'}_f\cdot{\bf i'},\,{\rm and}\,
\dot\psi_{fy'}=\dot\psi_f{\bf k'}_f\cdot{\bf j'}$, we can write
\begin{eqnarray}
\dot\psi_{fx'}&=&\dot\psi_f\left(\frac{p_fq_m}{\sqrt{p_m^2+q_m^2}}-
\frac{q_fp_m}{\sqrt{p_m^2+q_m^2}}\right)\nonumber\\
 \dot\psi_{fy'}&=&\dot\psi_f\left(-\frac{p_fp_m\sqrt{1-p_m^2-q_m^2}}
{\sqrt{p_m^2+q_m^2}}-\frac{q_fq_m\sqrt{1-p_m^2-q_m^2}}
{\sqrt{p_m^2+q_m^2}}+\sqrt{(1-p_f^2-q_f^2)(p_m^2+q_m^2)}\right)\nonumber\\
\dot\psi_{fz'}&=&\dot\psi_f\left[p_fp_m+q_fq_m+\sqrt{(1-p_f^2+
q_f^2)(1-p_m^2-q_m^2)}\right].\label{eq:psicdot} 
\end{eqnarray}
Substitution of Eqs. (\ref{eq:psicdot}) and (\ref{eq:bfibfj}) into
Eq. (\ref{eq:presstorque2}) yields ${\bf T}_{\s P}$ in terms of the
variables of Eq. (\ref{eq:finaleqs}). 

This completes the development of the appropriate torques acting on
Mercury's mantle and core. These torques are substituted into
the components of ${\bf N_m}$ and ${\bf N_f}$ in
Eqs.(\ref{eq:finaleqs}).  

\section{Results\label{sec:results}}

\begin{figure}[h]
\begin{center}
\epsscale{1}
\plottwo{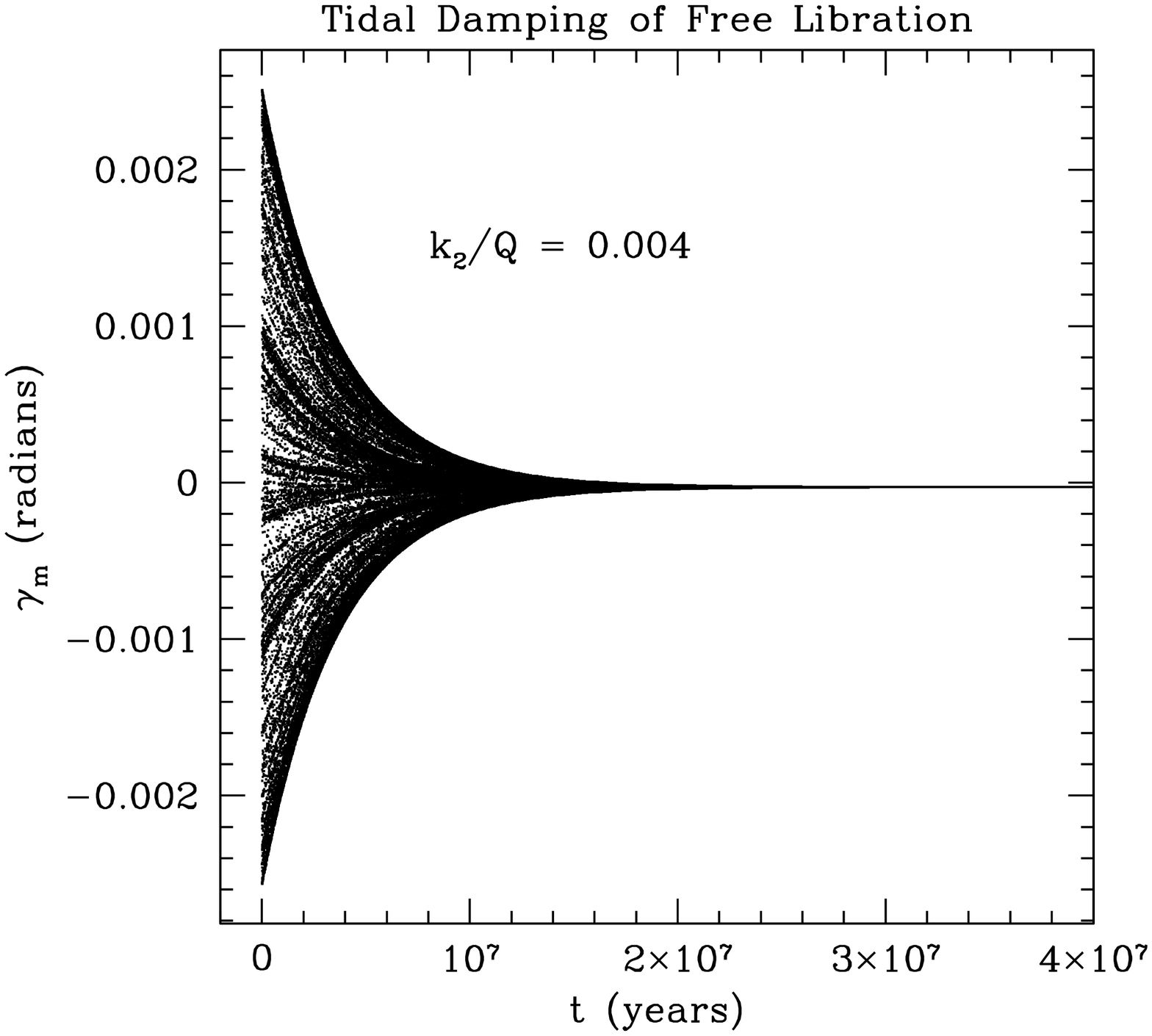}{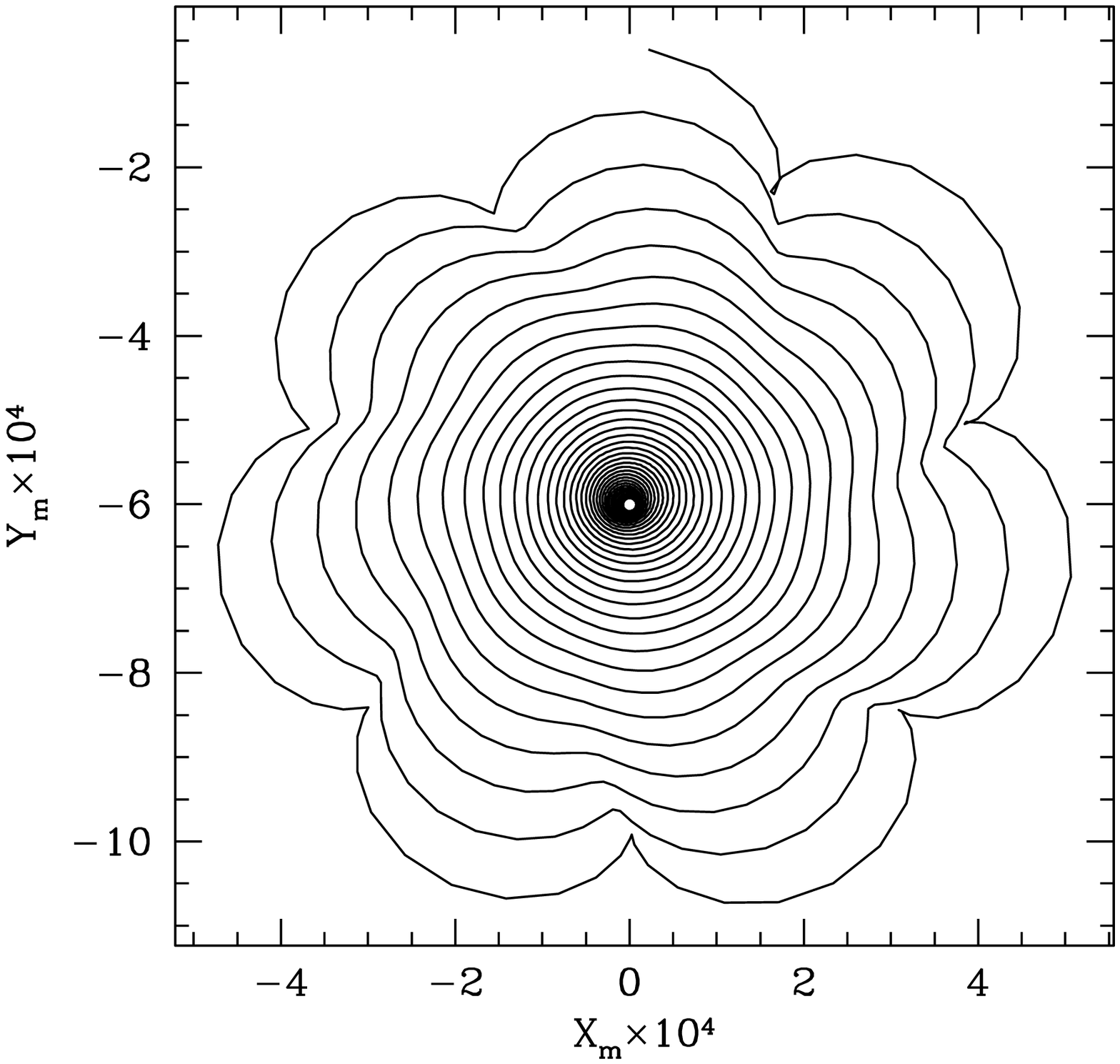}
\caption{Example of damping the initial free libration for tidal
torques alone (left panel) and the precession amplitude of the mantle
for pressure torque and viscous torque (right panel), where
$k_2/Q=0.004$ for the left panel and viscosity $\nu=2.25\times 10^5
(1.0)\,{\rm cm^2/s}$ 
for the right panel. $X_m$ and $Y_m$ indicate the position of the
unit vector along the mantle spin axis projected onto the $XY$ orbit
plane. Since the angles are small, the coordinates represent
radians ($6\times 10^{-4}\,{\rm rad}=2.06$ arcmin).   
Initial conditions: $(i_m,\Omega_m)=(0.1^\circ,0^\circ)$, $(i_f,
\Omega_f)=(0.1^\circ,180^\circ)$, $\gamma_m=0.1^\circ$,
$\dot{\gamma}_m=0.0001^\circ/{\rm day}$, and 
$\dot{\gamma}_f=0.0001^\circ/{\rm day}$. \label{fig:evolution}}. 
\end{center}
\end{figure}

We  apply the dissipative torques separately to determine the effect
of each for a spherically symmetric CMB.  In each case, the final
state is determined by running the calculations until free librations
in longitude are completely damped, and 
there is no longer any change in a very small circulation of the spin
vector around the mean position of the equilibrium state. We calculate
the evolution for the tides alone for a solid planet whose moment of
inertia $C/mR^2=0.346$ (Margot et al. 2012). We are interested in the
final evolutionary state of Mercury's spin, so we start the system
close to that state. Fig. \ref{fig:evolution} shows an example of the
damping of the free librations in longitude for tidal evolution alone with
$k_2/Q=0.004$ (Van Hoolst and Jacobs (2003) calculate values of
$k_2\sim 0.37\,{\rm to}\,0.6$ for a range of core sulfur content and
inner core radii. The value of $k_2/Q$ chosen is for a typical $Q$
value near 100.) The second panel in Fig. \ref{fig:evolution} shows
the damping of the precessional amplitude with viscous and pressure
torques applied. The general behavior of the system for
any of the individual core--mantle torques is similar to that displayed
in Fig. \ref{fig:evolution}, although the final equilibrium positions 
as well as the time scales to reach equilibrium differ.  For example,
for the tides alone the time scale to reach equilibrium exceeds
20 million years, whereas that for the viscous torque 
with the parameters for the right panel of
Fig. \ref{fig:evolution} is on the order of a million years.  

\subsection{Tidal evolution}

Tidal evolution for a solid Mercury essentially takes the system 
to the Cassini state, as shown in the left panel of
Fig. \ref{fig:viscosity}. The Cassini state is in fact defined for the
solid planet with the full moment of inertia. The final equilibrium
states for viscous core--mantle coupling are also shown in
Fig. \ref{fig:viscosity} for several values of the kinematic viscosity
$\nu$. The small circles in this figure traced by the projection of
the spin onto the XY plane are the endpoints of the evolution; the
small circles result from the precession of the perihelion of
Mercury's orbit. The projection of the spin axis onto the orbit plane
makes one revolution around a small circle as the perihelion argument
$\omega$ makes one half 
revolution.  The radii of the circles are less than the maximal 1
arcsec fluctuations in the position of spin relative to the
instantaneous Cassini state because of variations in the solar system
parameters that determine the state (Peale, 2006).  Peale (1974)
showed that the definition of the Cassini state as being coplanar with
the orbit and Laplace plane normals is a good approximation, where the
argument of perihelion is absorbed into the libration angle. The small  
scatter of the equilibrium positions of the spin for different
values of $\omega$ shown in Fig. \ref{fig:viscosity} displays the
deviations from the approximate solution defining the state; these
deviations are considerably smaller than the one standard deviation
($1\sigma\sim 5$ arcsec) observational uncertainty in the position
of the spin axis indicated by the small ellipse (Margot {\it et al.},
2012).    

When any of the dissipative CMB torques is turned on, it completely
dominates the tidal torque on the mantle, except in the limit of very
small coupling.  In this limit, both the tidal torque and the small
CMB torque drive the mantle spin to another Cassini state appropriate
to the moment of inertia of the mantle alone ($C_m/mR^2=0.149$) at a
smaller obliquity. For stronger core--mantle coupling, the final
equilibrium states of the mantle spin for all the CMB torques are not
distinguishably altered by the tidal torques.

\subsection{Viscous coupling}

\begin{figure}[h]
\begin{center}
\epsscale{1}
\plottwo{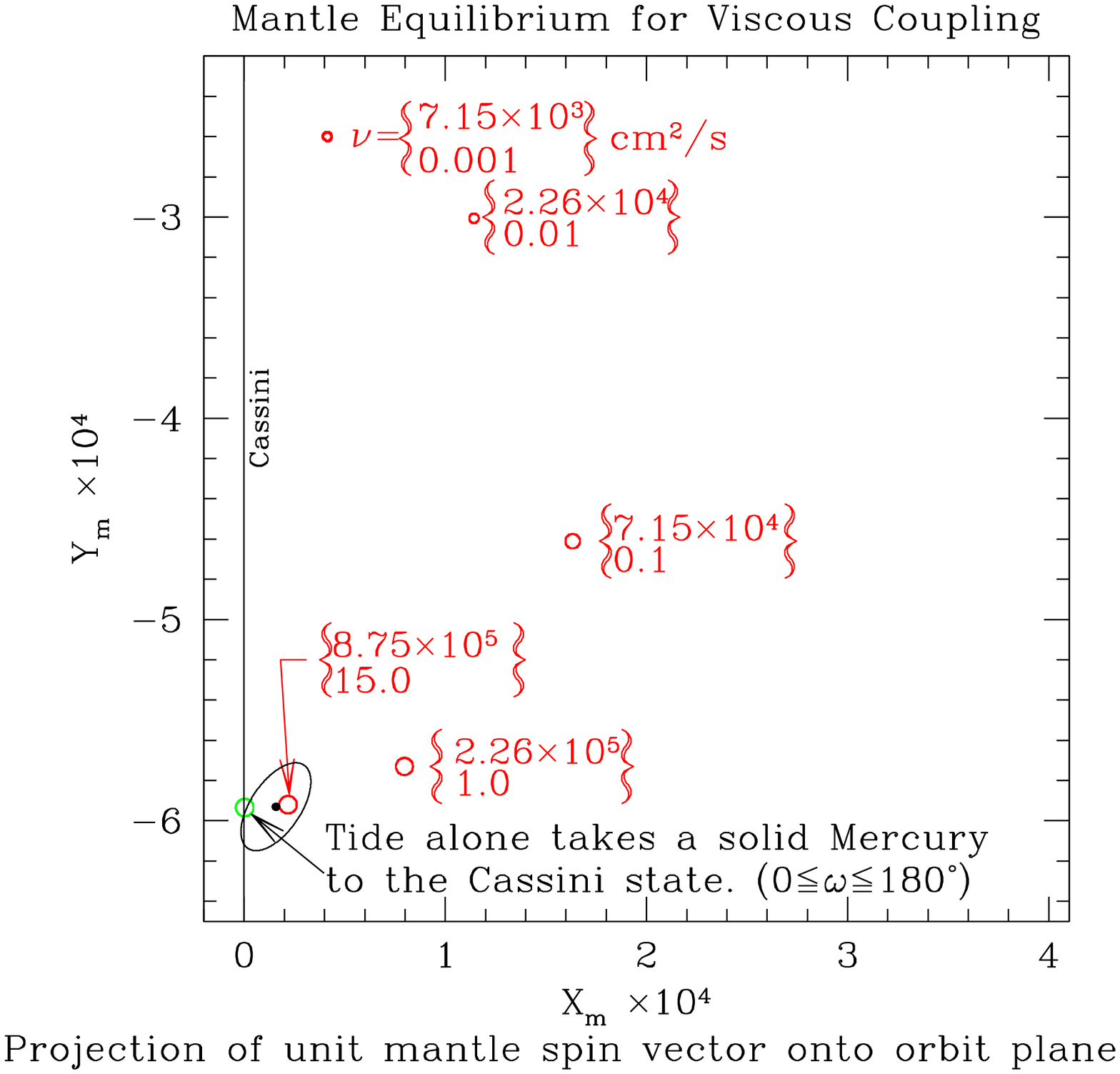}{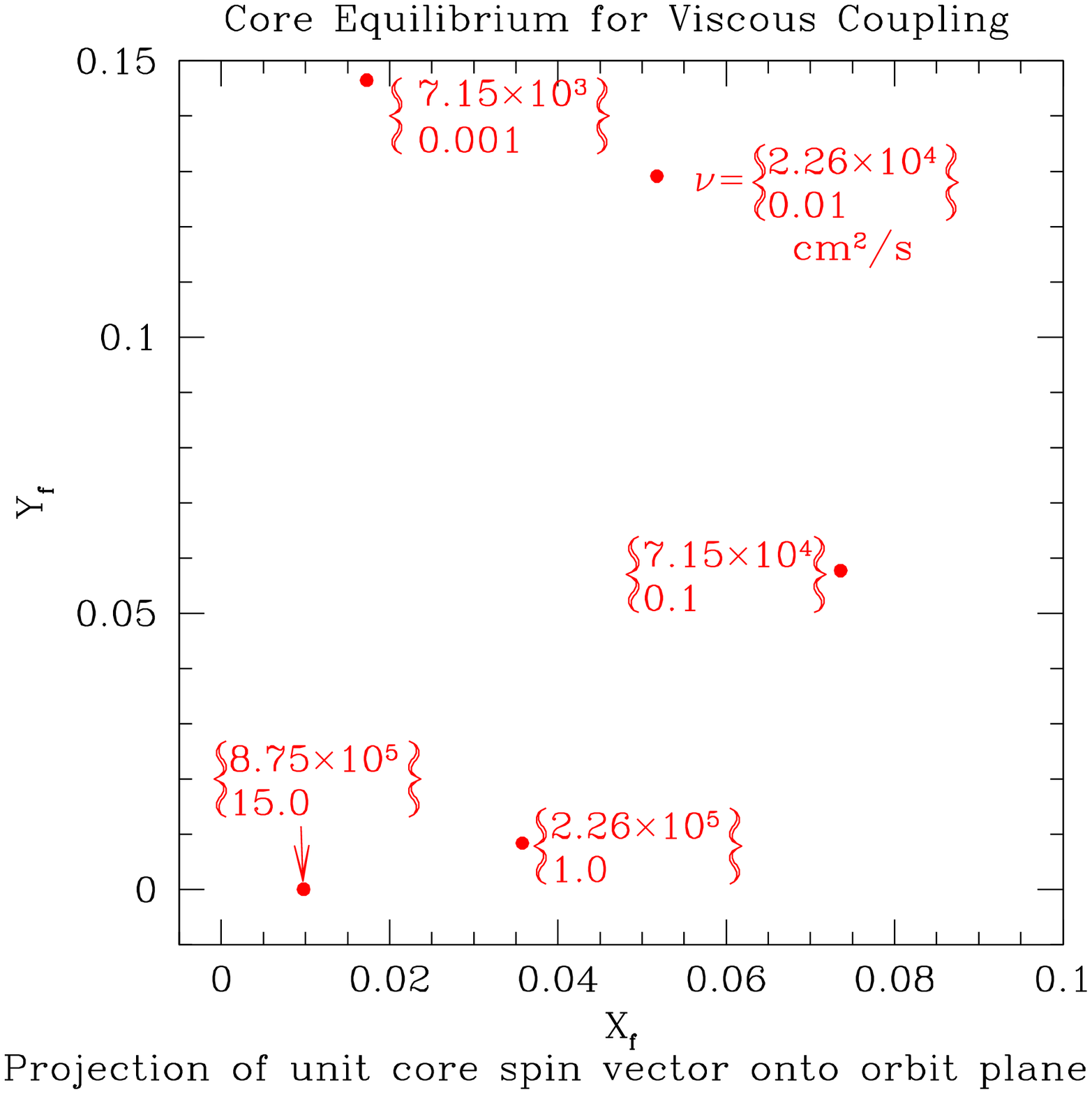}
\caption{Equilibrium offset of the mantle spin axis from the Cassini
state and corresponding positions of the core spin axis for two types
of viscous coupling and several values of the viscosity (Table
\ref{tab:viscous}). 
The pairs of viscosities at each point refer to the same value of the
coupling constant $\beta$ for viscosity time scales
$R_f/(\nu\dot\psi)^{1/2}$ (smaller viscosity) and $R_f^2/\nu$ (larger
viscosity) discussed in the text. A viscosity $\sim 8.75\times
10^5\,({\rm or 15})\,{\rm cm^2/s}$ is necessary to bring the spin
axis to within the $1\sigma$ uncertainty (small ellipse) in the pole
position near the 
Cassini state for the two time scales. In the left panel the small
circles describe the variation of the final equilibrium position as 
the perihelion longitude varies. The equilibrium state of the
mantle  for tidal evolution of a completely solid Mercury is also
shown by the small circle on the line labeled Cassini, which is the
intersection of the plane containing the Cassini state (spin vector),
the orbit normal (at $X_m,Y_m=0,\,0$), and the Laplace plane normal (at
$X_m,Y_m=0,\,0.15$) with the orbit plane. The best observational
determination of the spin position is the dot in the center of the
$1\sigma$ uncertainty ellipse.\label{fig:viscosity}}      
\end{center}
\end{figure}

The final equilibrium states of the system for viscous coupling
between the core and mantle displayed in Table \ref{tab:viscous} and
Fig. \ref{fig:viscosity} show large offsets of the spin axis from the
Cassini state.  Each viscosity in a pair for the two
relaxation time scales yields the same value of $\beta$
(Eqs. (\ref{eq:beta1}) and (\ref{eq:beta2})).  The relatively
short time scale for approaching the equilibrium state indicated in
the right panel of Fig. \ref{fig:evolution} is typical of the
viscous coupling as well as the magnetic and topographic coupling
considered below.  The offset is
largest for the smallest viscosities, and one
must increase the core--mantle coupling to bring the spin axis to within
the uncertainty of the observed position.  The kinematic viscosity
$\nu\sim 15\,{\rm cm^2/s}$ or $\nu\sim 8.75\times 10^5\,{\rm cm^2/s}$
for the two 
relaxation time scales for this condition to be satisfied. Binding
the core more strongly to the mantle means that Mercury approaches the
motion of a solid planet, and hence the mantle will end up closer to
the Cassini state so defined. 
\begin{table}[h]
\caption{ Equilibrium positions (projections) ($X_i,Y_i$)  and phase of
mantle ($\gamma_m$) and deviation of core angular velocity from $1.5n$
($\dot\gamma_f/n$) for viscous coupling
(Fig. \ref{fig:viscosity}). The equilibrium angular 
velocity of the mantle is $\equiv 1.5n$. The last  line shows the
equilibrium position for tidal evolution of a completely solid Mercury.} 
\label{tab:viscous} 
\centering
\begin{tabular}{ccccccc}
\hline
$\nu\,{\rm cm^2/s}$&$X_m\times 10^4$&$Y_m\times 10^4$&$X_f\times
10^2$&$Y_f\times 10^2$&$\gamma_m\times 10^4$&$\dot{\gamma_f}/n\times 10^3$\\
\hline
\hline
0.001, $7.15\times 10^3$&0.4050&-2.6176&1.7282&0.1466&-0.2979&-14.6552\\
0.01, $2.26\times10^4$&1.1600&-3.0221&5.1764&12.9136&-0.5569&-12.9136\\
0.1, $7.15\times 10^4$&1.6421&-4.6545&7.3584&5.7722&-0.6586&-6.5967\\
1.0, $2.26\times 10^5$&0.8275&-5.7003&3.5782&0.8374&-0.4675&-0.8374\\
15.0, $8.75\times 10^5$&0.2576&-5.9385&0.9787&0.0037&-0.3346&-0.0373\\
Tide&0.0&-5.9256&&&-2.8457&\\
\hline
\end{tabular}
\end{table}

\begin{figure}[h]
\begin{center} 
\epsscale{.6}
\plotone{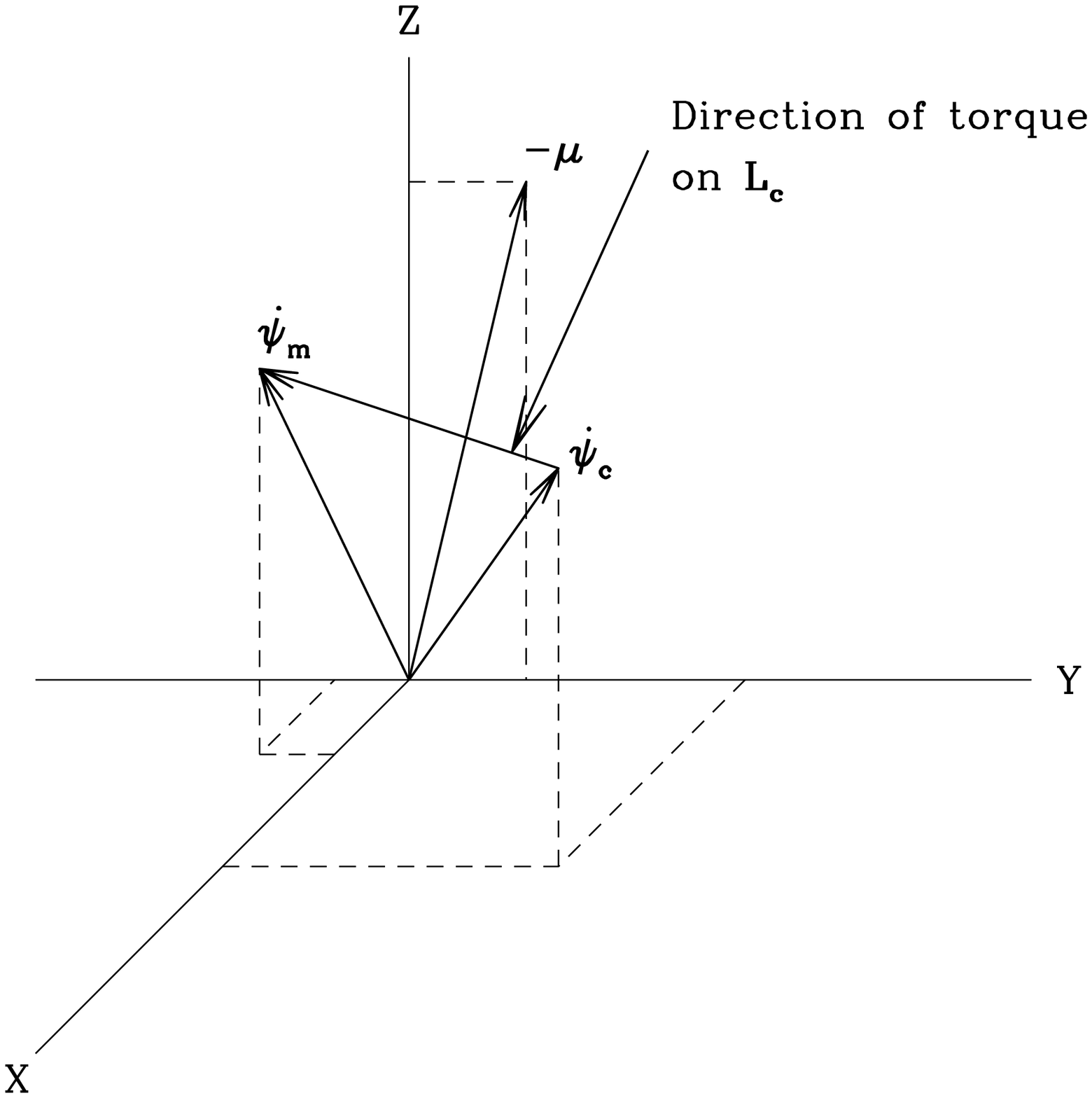}
\caption{Schematic of the final positions of the unit spin vectors of
the core and mantle for one 
of the smaller viscosities in Fig
\ref{fig:viscosity}. The torque on the core causes the precise
precession of the core with the orbit. \label{fig:equilib}}
\end{center}
\end{figure}

The large differential rotation  between the mantle and core induced
by the orbit precession is initially surprising.  The mantle spin is
projected  onto the fourth quadrant of the XY orbit plane, whereas the
core spin is projected onto the first quadrant, which is shown
schematically in Fig. \ref{fig:equilib}. The direction of the viscous
torque is proportional to $\bs{\dot\psi}_m-\bs{\dot\psi}_f$.  That this 
torque is perpendicular to the precessional angular velocity can be
inferred from the figure.  An evaluation  of all the torques on the
core and mantle when the system has reached the equilibrium position
confirms the simultaneous precession of both with the precession of
the orbit, as they must to be stationary in the orbit frame of
reference.  What has happened is that 
the core and mantle have assumed positions such that the total torque
on each is just that necessary to cause each to precess with the
orbit. When the coupling is weak, the vector separation of the spins of
mantle and core is increased to provide the necessary torques, which
in turn leaves the mantle more separated from the Cassini state.  
The equilibrium values of the
parameters for viscous coupling are shown in Table \ref{tab:viscous} for
$\omega=0$. Notice that the 
core rotation about the spin axis is slightly below the spin--orbit
resonant value of $1.5n$. The mantle always ends up with
$\dot\gamma_m\equiv 0$, and the latter is omitted from Table
\ref{tab:viscous}. 

The kinematic viscosity of the Earth's core has a wide range of estimates.
From damping of nutational motions, Smylie {\it et al.} (2009) determined
a value near the CMB close to $\nu\sim 2900\,{\rm cm^2/s}$. But from
laboratory measurements and theoretical simulations that limit
the variation of viscosity of liquid iron with pressure, $\nu\sim
10^{-2}\,{\rm cm^2/s}$  (de Wijs {\it et al.} 1998). Vo\v{c}adlo
(2007) discussed the 14 orders of magnitude range in the estimates of
Earth's core viscosity, but she favored the laboratory and theoretical
estimates that place the value of $\nu$ near that of  water at
Earth's surface.  All 
of the group of small viscosities in Fig. \ref{fig:viscosity} fall
within the smaller range spanned by the two estimates by    de Wijs {\it
et al.} (1998), but the
group of large viscosities all exceed the larger value.  If
Mercury has a solid layer of FeS at the base of the mantle that
separated from an Fe--S--Si mixture in the core (Hauck {\it et al.},
2013), the fluid adjacent to the CMB might be a slurry of solid FeS
particles with a viscosity comparable to or even exceeding the larger
viscosities in Fig. \ref{fig:viscosity}. There remains almost an
unconstrained uncertainty in the core viscosity.

Since the differential rotation of the core and mantle is relatively
large (up to several degrees separating the spin vectors), it is
prudent to check the energy dissipation resulting from 
the differential motion.  The rate of work being done by the viscous
core--mantle torque on the core is just the scalar product of this
torque and the differential angular velocity.  Hence,
\begin{equation}
\frac{dW}{dt}=\beta(\bs{\dot\psi}_m-\bs{\dot\psi}_f)\cdot
(\bs{\dot\psi}_m-\bs{\dot\psi}_f).\label{eq:dWdt} 
\end{equation}
The components of $\bs{\dot\psi}_m$ and $\bs{\dot\psi}_f$ are given
in Table \ref{tab:viscous}; the scalar magnitudes 
are
$\dot\psi_m=1.5n$ and 
$\dot\psi_f=(1.5n+\dot\gamma_f)$.  From Eqs. (\ref{eq:beta1}) and
(\ref{eq:beta2})  we find 
$\beta=9.307\times 10^{29}\,{\rm and}\,3.605\times
10^{31}\,{\rm g\,cm^2/s}$, respectively, for $\nu=2.26\times
10^4,\,0.01{\rm and}\,8.75\times 10^5,\,15\,{\rm cm^2/s}$. Also,
$(\bs{\dot\psi}_m-\bs{\dot\psi}_f)^2=3.809\times 10^{-2}n^2\,{\rm
and}\,2.188\times 10^{-4}n^2$ for the two extremes, respectively.  Then
\begin{eqnarray}
\frac{dW}{dt}&=&2.423\times 10^{16}\,{\rm
erg/s},\;\;\;\;\nu=(2.26\times 10^4,\,0.01)\,{\rm cm^2/s},\nonumber\\
&=&1.213\times 10^{16},\,{\rm erg/s}\;\;\;\;\nu=(8.75\times
10^5,\,15.0)\,{\rm cm^2/s}. 
\label{eq:dWdt2}
\end{eqnarray}
The rate at which the torque does work is smaller in spite of the
larger viscosity because the relative motion of the core and mantle is
so much smaller that it overcompensates the larger viscosity.   

The consequence of the torque on the core is to make the core precess
with the orbit (core spin fixed in the orbit frame). The work done on
the core by the torque is an upper bound on the energy dissipated.
To judge the importance of this dissipation, we compare it with the
radioactive heat production in the mantle and with the rate at which
energy is conducted into the mantle at the CMB. 

The current radioactive heat production in the Earth's mantle is
$\sim 7.4\times 10^{-8}\,{\rm erg/(g\,s)}$ from $^{238}{\rm
U}$, $^{235}{\rm U}$, $^{232}{\rm Th}$, and $^{40}{\rm K}$ (Turcotte
and Schubert, 2002). Peplowski {\it et al.} (2011) find that the
current radioactive heat generation from K, Th, and U on Mercury's
surface is near $2\times 10^{-7}\,{\rm erg/(g\,s)}$. Michel {\it et al}
(2013) and Tosi {\it et al.} (2013) model Mercury's thermal history
constrained by present day surface heat production, global contraction
estimates, and time frames of magma production. Both models are
consistent with about a factor of three enhancement in the
concentration of radioactive elements on the surface compared with the
current mantle concentration, which leads to a current mantle heat
production near $7\times 10^{-8}\,{\rm erg/(g\,s)}$---comparable to
that in the Earth's mantle. For a core radius of 2000 km, a mean
Mercury radius of 2440 km, and a mean mantle density near $3.2\,{\rm
g/cm^3}$ (Table {\ref{tab:alphabeta}}), the total radioactive heat  
production in the mantle is near $6.5\times 10^{18}\,{\rm erg/s}$
with no contribution from the crust. 

Alternatively, for the conduction of heat into the mantle at the CMB, the 
thermal conductivity of FeS is $K_1=3.5\times 10^5\,{\rm
erg/(s\,cm\,K)}$ at $308\,{\rm K}$ (Table 3 of Clauser and Huenges,
1995). There are few data on the temperature and pressure dependence
of thermal conductivity of FeS, but that for single crystal ${\rm
FeS_2}$ varies inversely with temperature between 200 and 300 K (Popov
{\it et al.}, 2013).  If we assume a similar dependence for FeS, the
thermal conductivity $K_1$ could be as small as $7\times 10^4\,{\rm
erg/(s\,cm\,K)}$ at 1500 K. This low value could be compensated
somewhat by an 
increase in the conductivity due to pressure ({\it e.g.}, Clauser and
Huenges, 1995) at the depth of the FeS layer, so we choose $K_1=1\times
10^5 {\rm erg/(s\,cm\,K)}$ for the FeS layer. The thermal conductivity
of basalt is near $K_2=2\times 10^5\,{\rm erg/(s\,cm\,K)}$ for a wide
range of temperature 
and pressure (Figs. 2 and 3 in Clauser and Huenges, 1995).    

With $R_f, R_{FeS}$,
and R the radii of 
the core, FeS layer, and the planet, we can write for a steady
state flux of heat $F$
\begin{eqnarray}
-\frac{dT}{dr}&=&\frac{F}{4\pi K_1 r^2}\;\;\;\;R_f<r<R_{FeS}\nonumber\\
-\frac{dT}{dr}&=&\frac{F}{4\pi K_2 r^2}\;\;\;\;R_{FeS}<r<R.\label{eq:dTdr}
\end{eqnarray}
Integration of these equations over the range of radii where each is applicable
yields two equations that give the difference in temperatures $T_{\s
R}$ and $T_{\s FeS}$ between $R$ and $R_{FeS}$ and separately $T_{\s
FeS}$ and $T_{\s R_f}$ between $R_{FeS}$ and $R_f$. Eliminating
$T_{FeS}$ from these equations, we can solve for the flux $F$ in terms
of $T_{\s R_f}-T_{\s R},\, K_1,\,K_2,\,R,\,R_{\s FeS},\,{\rm
and}\,R_f$. With $F=-4\pi R_f^2K_1dT/dr$ at $R_f$, we can solve for
the temperature gradient at the CMB appropriate to the conductivities
and a given temperature difference between a layer close to the surface
and the CMB.  
\begin{equation}
\frac{dT}{dr}\Bigg |_{R_f}=\frac{\displaystyle T_{\s R_f}-T_{\s
R}}{\displaystyle R_f\left[\frac{R_f}{R_{\s FeS}}-1+\frac{K_1}{K_2}\left(
\frac{R_f}{R}-\frac{R_f}{R_{\s FeS}}\right)\right]}. \label{dTdr}
\end{equation}
Solutions of the interior structure show that it is statistically
unlikely for the thickness of a solid FeS layer at the base of the
mantle to exceed 90 km (Hauck {\it et al.}, 2013).   
With $R_f=2040\,{\rm km}$, $R_{\s FeS}=2130\,{\rm km}$,
$R=2440\,{\rm km}$, $T_{\s R_f}=1500\,{\rm K}$ (near the melting point
of FeS), $T_{\s R}=300\,{\rm K}$ (as an average temperature over a
surface a short distance below the surface of Mercury), and the above
values of the conductivity coefficients, we find $dT/dr|_{\s
R_f}=5.706\times 10^{-5}\,{\rm K/cm}$.  

Substitution of this gradient back into the expression for $F$ yields
\begin{equation}
F=2.98\times 10^{18}\,{\rm erg/s}. \label{eq:thermalflux}
\end{equation}
The maximum viscous dissipation at the CMB of $2.423\times
10^{16}$ or $1.213\times 10^{16}\,{\rm erg/s}$ for the extremes in
the viscosity pairs is less than a 1\% contribution to either the
radioactive heat production in the mantle or the
heat flux from normal thermal conduction at the CMB, and should thus
not alter the temperature distribution or any observable quantity
significantly. Below we show that pressure coupling restricts the
differential rotation between core and mantle to much smaller values,
so viscous dissipation will contribute even less to the total heat
flux. 

\subsection{Magnetic coupling}

\begin{figure}[h]
\begin{center}
\epsscale{1}
\plottwo{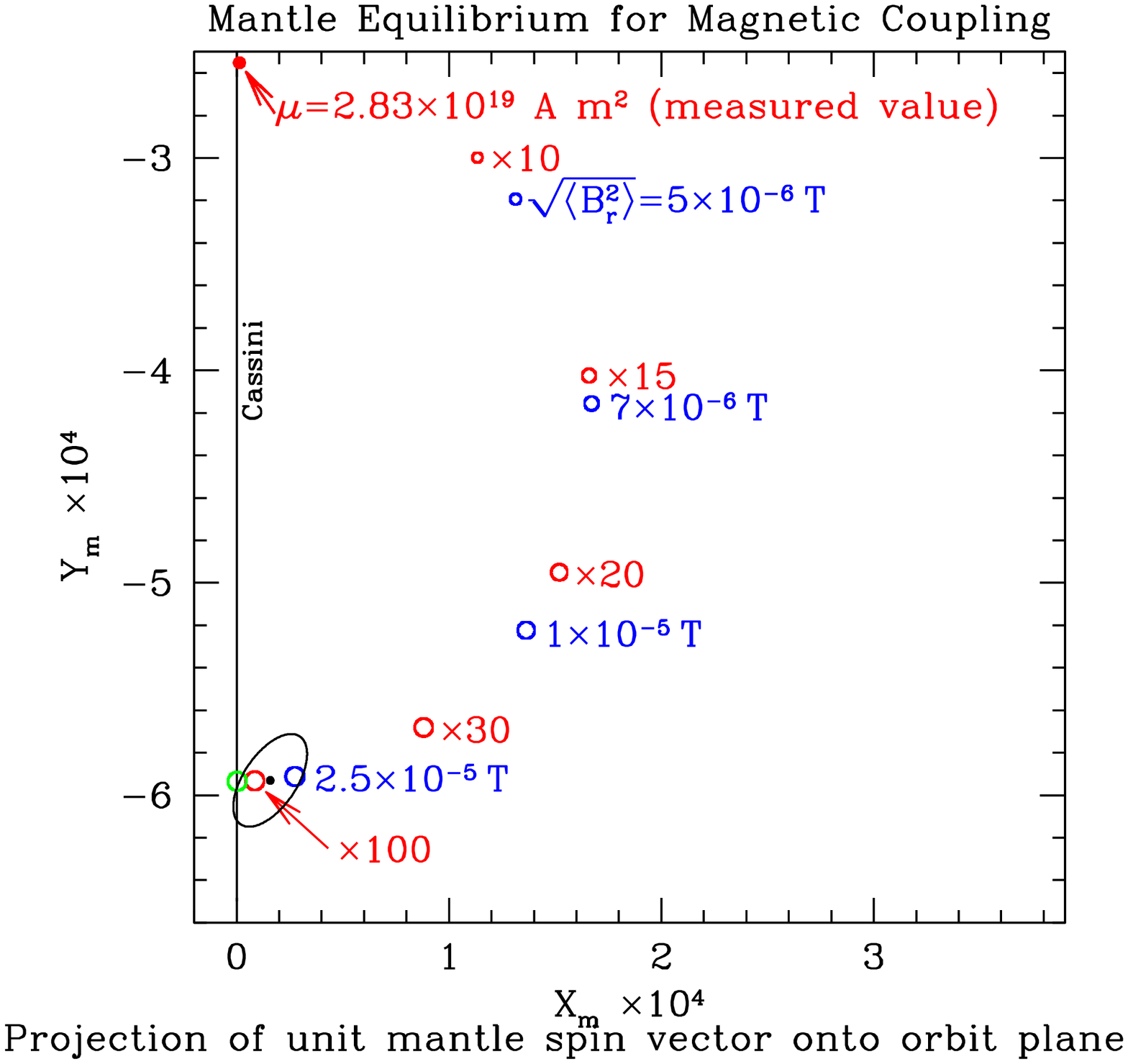}{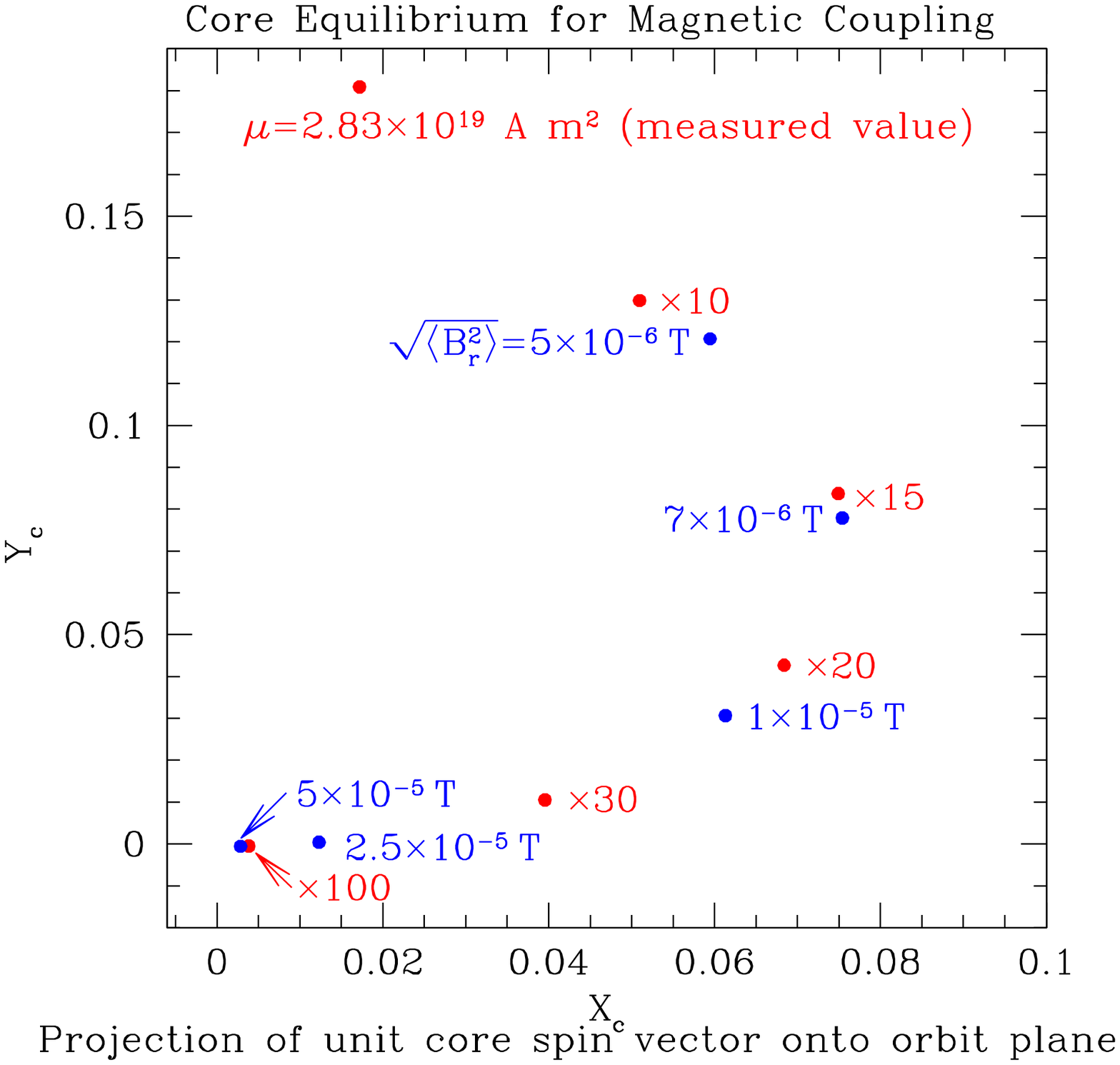}
\caption{Equilibrium offset of the mantle spin axis from the Cassini
state and corresponding positions of the core spin axis for magnetic
core--mantle coupling for several values of the magnetic 
dipole moment $\mu$ and separately for an rms value of $B_r$ on the
CMB for dipole plus higher-order terms (Table \ref{tab:magnetic}).  For
dipole coupling alone, a 
magnetic moment that is 100 times the value of 
$2.38\times 10^{19}\,{\rm A\,m^2}$ inferred from the magnetic field
measurements is necessary to bring the
spin of the mantle to within the uncertainty in the pole
position. Alternatively, an rms value of $B_r\gtwid 2.5\times
10^{-5}\, {\rm T}$ is required. Other details in this figure are as in
Fig. \ref{fig:viscosity}.
 \label{fig:magnetic}}
\end{center}
\end{figure}

Magnetic dipole coupling leads to behavior similar to that of the
viscous coupling with the offset of the mantle spin axis from the
Cassini state being quite large for the observed value of the magnetic
moment of $ 2.43\times10^{19}\,{\rm A\,m^2}$ (Fig. \ref{fig:magnetic}). The
equilibrium positions of the mantle spin progress in an arc toward the
Cassini state as the coupling is increased by increasing the magnetic
dipole moment.  This coupling is for a centered dipole, whereas
Mercury's dipole is offset to the north by about 486 km (Anderson
{\it et al.},  2011, 2012). However,
since the coupling is so small, it was thought unnecessary to recover no
more than a factor of a few in the coupling magnitude
with the offset dipole. Instead, we determine
the equilibrium positions of the spin axes for a root mean square
(rms) value of the 
radial component of the field $B_r$ on the CMB. A value of the
magnetic dipole moment necessary to bring the spin axis close to the
Cassini state is a factor of 100 more than the value inferred from the
external field measurements. The
rms value of $B_r\,{\rm is}\,2.5\times 10^{-5}\,{\rm T}$ for the same
condition. For comparison, the rms value of $B_r\,{\rm is}\,2.25\times
10^{-7}\,{\rm T}$ on the CMB for the observed dipole field. The
parameter values at the equilibrium state for 
magnetic coupling are given in Table \ref{tab:magnetic} for $\omega=0$.
Magnetic coupling between the core and mantle is clearly not
sufficient to bring the spin axis of the mantle to within the
observational uncertainty of its position for magnetic field strengths
at the CMB inferred from measurements in the magnetosphere.

\begin{table}[h]
\caption{Equilibrium positions (projections) ($X_i,Y_i$) and phase of the
mantle ($\gamma_m$)  and the deviation of core angular velocity from
$1.5n$ ($\dot\gamma_f/n$) for magnetic coupling
(Fig. \ref{fig:magnetic}). The upper 
half of the table is for dipole coupling only with the first column
being the factor multiplying the measured magnetic moment for Mercury
of $2.83\times 10^{19}\,{\rm A\,m^2}$. The lower half of the
table gives equilibrium positions for rms values of $B_r$ on the CMB
from all harmonic components of the field.} 
\label{tab:magnetic}
\centering
\begin{tabular}{ccccccc} 
\hline
Moment&$X_m\times 10^4$&$Y_m\times 10^4$&$X_f\times
10^2$&$Y_f\times 10^2$&$\gamma_m\times 10^4$&$\dot{\gamma_f}/n\times 10^3$\\
\hline
\hline
1.0&0.0961&-2.7593&0.0065&14.9464&-0.0119&-41.7039\\
$10\times$&1.0735&-3.1859&5.0955&12.9856&-0.2487&-43.1722\\
$15\times$&1.5921&-4.1803&7.4812&8.4525&-0.3688&-28.4403\\
$20\times$&1.4759&-5.1006&6.8718&4.3378&-0.3419&-14.8230\\
$30\times$&0.8639&-5.8424&3.9929&1.0750&-2.0014&-3.8547\\
$100\times$&0.0843&-6.1011&0.3885&0.0511&-0.0192&-0.0384\\
\hline
\hline
$\sqrt{\langle B_r^2\rangle}$&$X_m\times 10^4$&$Y_m\times 10^4$&$X_f\times
10^2$&$Y_f\times 10^2$&$\gamma_m\times 10^4$&$\dot{\gamma_f}/n\times 10^3$\\
\hline
$5\times 10^{-6}\,{\rm T}$&1.3538&-3.4892&6.2491&11.6587&-0.3232&-13.2290\\
$7\times 10^{-6}\,{\rm T}$&1.6396&-4.4886&7.5437&7.1573&-0.3771&-8.1619\\
$1\times 10^{-5}\,{\rm T}$&1.2703&-5.4904&5.8239&2.6697&-0.2921&-3.0930\\
$2.5\times 10^{-5}\,{\rm T}$&0.2474&-6.0838&1.1325&0.02394&-0.0569&-0.0963\\
$5\times 10^{-5}\,{\rm T}$&0.0634&-6.1016&0.2846&-0.0557&-0.0143&-0.00605\\
\end{tabular}
\end{table}

Some insight into what is happening for equilibrium, when the spin axes
of both the core and mantle are stationary in the precessing orbit
frame, can be found by examining the magnetic dipole coupling
positions in Fig. \ref{fig:magnetic}.  For the observed magnetic
dipole moment, the spin axis is nearly in the plane defined by the
Laplace plane normal and the orbit plane normal.  This is again a
Cassini state, but for a planetary moment of inertia corresponding to
that of the mantle alone ($0.149mR^2$ instead of $0.346mR^2$).  In other
words, the coupling is so weak that the mantle rotates as if the core
were not there.  The similar behavior for the other core--mantle
coupling torques as the strength of the coupling is varied means they
would all approach this other Cassini state in the limit of weak
coupling. 

\subsection{Topographic coupling}

\begin{figure}[h]
\begin{center}
\epsscale{1}
\plottwo{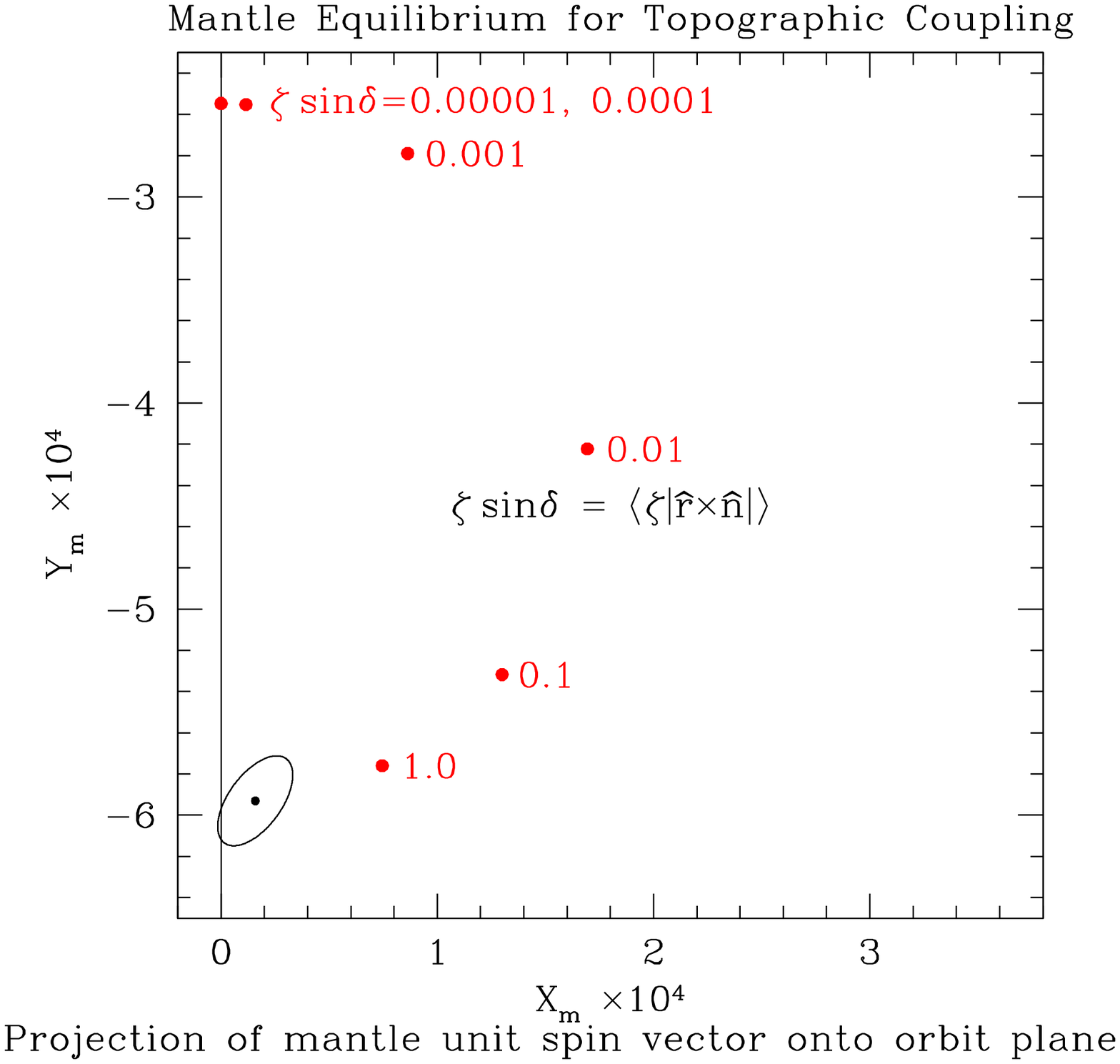}{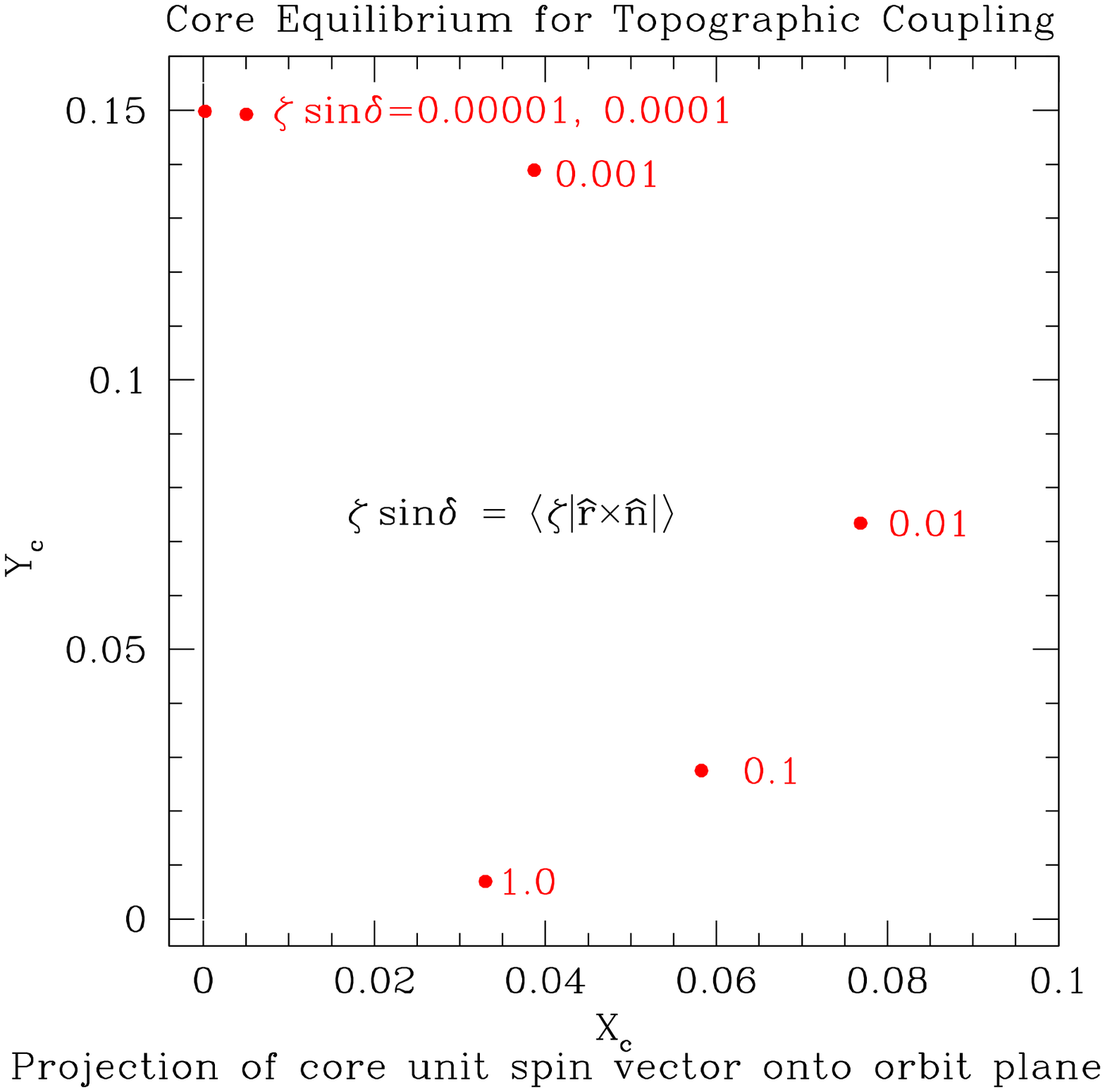}
\caption{Equilibrium offset of the mantle spin axis from the Cassini
state and corresponding positions of the core spin axis
for topographic core--mantle coupling for several values of slopes of
CMB bumps (Table \ref{tab:topographic}).  Even the maximum value of
$\zeta\sin{\delta}=1.0$ is 
insufficient to bring the mantle spin to within the uncertainty
ellipse locating the mantle spin position.  \label{fig:topography}}
\end{center}
\end{figure}

The distribution of equilibrium positions for the mantle and core for
topographic coupling is shown in Fig. \ref{fig:topography}. The
behavior is again similar to the other two core--mantle coupling
processes. Since the ad hoc efficiency $\zeta\le 1$ and
$\sin{\delta}\le 1$, the product cannot be sufficiently large to bring
the spin axis close to the observed value. 
So the modeled topography apparently cannot provide sufficient
coupling between core and mantle to this end. Table
\ref{tab:topographic} shows the 
parameter values appropriate to topographic core--mantle coupling.
\begin{table}[h]
\caption{Equilibrium positions (projections) ($X_i,Y_i$)  and phase
of the mantle ($\gamma_m$) and deviation of the core angular velocity
from 1.5n ($\dot\gamma_f/n$)  for topographic coupling
(Fig. \ref{fig:topography}).} 
\label{tab:topographic}  
\centering
\begin{tabular}{ccccccc}
\hline
$\zeta\sin{\delta}$&$X_m\times 10^4$&$Y_m\times 10^4$&$X_f\times
10^2$&$Y_f\times 10^2$&$\gamma_m\times 10^4$&$\dot{\gamma_f}/n\times 10^3$\\
\hline
\hline
0.00001&0.0&-2.5465&0.0200&14.9830&-0.0239&-0.1315\\
0.0001&0.1155&-2.5520&0.5034&14.9268&-0.0259&-0.1322\\
0.001&0.8627&-2.7890&3.8690&13.8924&-0.1979&-0.1097\\
0.01&1.6937&-4.2230&7.6860&7.3440&-0.3845&-0.0391\\
0.1&1.3000&-5.3174&5.8260&2.7530&-0.2992&-0.0083\\
1.0&0.8266&-5.7599&3.4400&0.6980&-0.1659&-0.0017\\
\hline
\end{tabular}
\end{table}

Like magnetic coupling, topographic coupling between mantle and core
also fails to bring the 
spin axis to within the observational uncertainty of its
position.  Only viscous coupling remains as 
a possibility. Even the latter is possible only for relatively large
values of the viscosity that may not prevail.  

\subsection{Pressure coupling \label{sec:pressure}}

\begin{figure}[h]
\begin{center}
\epsscale{.55}
\plotone{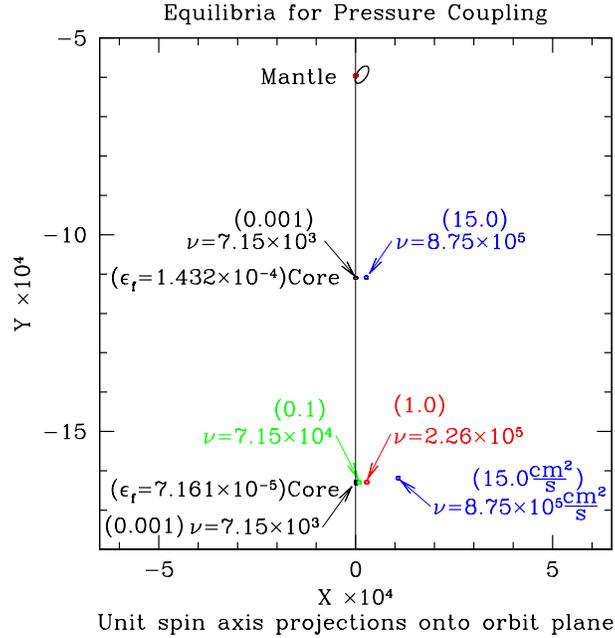}
\caption{Equilibrium positions of mantle and core spins for pressure and
viscous coupling at the CMB. $\nu$ values are for the viscous 
time scales; parentheses denote $\nu$ values for Greenspan and Howard
time scales. The values of $\epsilon_f$ appropriate for the two groups
of core positions are indicated in parentheses.  \label{fig:pressure}} 
\end{center}
\end{figure}

The ellipsoidal
shape of the CMB leads to pressure torques between core and mantle
that result in far different behavior from that of the dissipative
torques alone. With the pressure torque in effect, 
viscous dissipation now brings the mantle spin to the Cassini state for
any value of the viscosity and for either of the values of
$\epsilon_f=7.161\times 10^{-5}$ (hydrostatic) or $1.432\times 10^{-4}$
(non hydrostatic) for an axisymmetric CMB (Fig. \ref{fig:pressure}).
The equilibrium positions of the core spin for the two values of
$\epsilon_f$ are much closer to the mantle spin position than they were
for the dissipative processes alone, and they lie nearly in the plane
containing the Cassini state displaced away from the Laplace plane normal. 

We have included viscous dissipation as the most likely process
that can influence the spin positions of the core and mantle 
with pressure coupling, although tidal, magnetic, or topographic
dissipation will also work on longer time scales. In all cases shown in
Fig. \ref{fig:pressure} the mantle spin separation from the Cassini
state is immeasurably small. The largest displacement of the
mantle spin axis toward the
fourth quadrant is 0.055 arcsec for $\epsilon_f=7.161\times 10^{-5}$ and 
$\nu=8.75\times 10^5\;{\rm or}\;15\,{\rm cm^2/s}$. This value is reduced to
0.016 arcsec for the larger $\epsilon_f=1.432\times 10^{-4}$. The observational
uncertainty ellipse for 
the mantle spin is included in the figure. The core spin has joined
the mantle spin by being nearly on the Cassini state line and
separated from the mantle spin by 3.55 arcmin for the smaller
hydrostatic $\epsilon_f$ and by 1.77 arcmin for the larger $\epsilon_f$
in the direction away from the Laplace plane normal. Increasing the
CMB ellipticity forces the core spin to be closer to that of the
mantle. As the viscosity is increased, the core spin lags the
retrograde precession of the Cassini plane by greater amounts, with
the displacement near 23 arcsec for the hydrostatic $\epsilon_f$ and
kinematic viscosity $\nu=8.75\times 10^5\;{\rm or}\;15{\rm
cm^2/s}$ but lagging only 5.6 arcsec for the $\epsilon_f$ that is double
the hydrostatic value.  

The core spin displacement is consistent with the classical solution
of Poincar\'e (1910) for Earth's core, for which  
the consequence for the core flow of the mantle precession is a
uniform vorticity (consistent with our assumption of the core rotating
uniformly) with the vorticity vector displaced from the symmetry
(rotation) axis of the CMB by a small angle in the direction away from
the negative of the precession vector. Our displacements of core spin
from mantle spin of 3.55 and 1.77 arcmin are much larger than the O(1)
arcsec displacement that Melchior (1986) found for Earth's core. But
Earth is spinning more rapidly, and the precession rates are far
different.   The lag of the core spin behind the precessing Cassini
plane is consistent with the results of Stewartson and Roberts (1963),
Roberts and Stewartson (1965), and Busse (1968), who found a lag for
the spin of the Earth's core in the same direction for the effect of
viscous dissipation. Here our final state for Mercury is obtained by
dissipative evolution starting from an arbitrary initial state. 
 
\section{Summary and conclusions\label{sec:summary}}

We have calculated the evolution of Mercury's mantle spin axis under
conservative and dissipative torques.  Tidal torque alone brings a solid
Mercury to Cassini state 1 with only a negligibly small offset of the
mantle spin axis from the Cassini state.  But each of the three
dissipative core--mantle interactions, treated alone, can lead to equilibrium
positions of the spin axis that are far outside the uncertainties in
its observed position.  For viscous coupling, a kinematic viscosity
$\nu\sim 8.75\times 10^5\;{\rm or}\;15 {\rm cm^2/s}$ is needed to
bring the spin axis within the $1\sigma$ 
uncertainty in the observed position. For magnetic dipole coupling,
the magnetic moment must be 100 times larger than that observed by the
MESSENGER spacecraft to satisfy the same constraint.  The rms value of
the radial component of the magnetic field $\sqrt{B_r^2}$  on the CMB
that can include the offset dipole field and all the multipole fields
is $2.5\times 10^{-5}\,{\rm T}$, which is two orders of magnitude
greater than the rms value of the dipole field at the CMB and one
order of magnitude greater than the estimated rms radial field from
MESSENGER observations (C. L. Johnson, personal communication,
2012). The observed magnetic field does not provide sufficient coupling
between the core and mantle to be consistent with the observed
position of the pole.  Finally there appears to be no physically possible
slope of the topography on the CMB that can bring the mantle spin
axis to within the uncertainty in its observed position. 

The core maintains a differential rotation for all the dissipative
equilibrium positions, and the difference increases as the coupling
decreases. As the core--mantle coupling is increased to the point where
the mantle spin axis equilibrium approaches the Cassini state, the
core spin approaches the orbit normal. This state can be only
temporary if the coupling continues to increase, as eventually the
core would be rigidly attached to the mantle, the spins would be
parallel and would occupy the Cassini state for a completely solid
Mercury, and the mantle would no longer librate independently of the
core.  

The ellipsoidal CMB leads to pressure coupling between the core and
mantle that completely controls the final equilibrium positions of the
spins, although the dissipative coupling is still necessary to effect the
evolution. The pressure torque in the presence of viscous (or any other)
dissipation drives the mantle spin to the Cassini state with 
negligible dissipative displacement. The final position of the core
spin is displaced by a few arcmin from the mantle spin approximately
in the Cassini plane in a direction away from the negative of the
precession vector. The 
viscous dissipation causes the core spin to lag the precession of the
Cassini plane by an amount that increases with the viscosity (23
arcsec for the largest viscosities shown in Fig \ref{fig:pressure}), a
configuration that is consistent with similar results 
for the Earth's core.  Unfortunately the close occupancy of the Cassini
state forced by the pressure coupling precludes using an offset of the
mantle spin to  constrain the dissipative properties of Mercury's
interior, and there is no obvious way to determine the core spin
orientation.  At the same time, we are assured of the theoretical
occupation of the Cassini state by Mercury's spin, which is so far
consistent with the observed position. This means that the constraints
on the interior structure that depend on knowledge of the Cassini
state obliquity are secure ({\it e.g.} Hauck, {\it et al.} 2013).
 
An unlikely caveat to this conclusion is the influence of a
non--spherical solid inner core.  This influence is suppressed by the 
pressure force on the inner core such that the effective density of a
deformed inner core is $\rho_s-\rho_f$ for its gravitational
interaction with the Sun and with the mantle.  If the light element
depressing the melting temperature of the fluid core is silicon
instead of sulfur, the density differential between the fluid and the
solid inner core is small (Hauck et al., 2013).  Still, it is of
interest to determine the behavior of the inner core for the sulfur
impurity and its possible influence on the equilibrium positions of
the observable mantle spin vector. Our conjecture is that this latter
influence is negligibly small, which is supported by the observed
position of the spin axis being consistent with occupancy of the
Cassini state, but this conjecture should be verified.  Since some of the
assumptions incorporated in this work, such as the uniform vorticity
of the fluid core found by Poincar\'e, may not be appropriate as the
thickness of the fluid core is reduced, we choose to investigate the
solid inner core behavior in a future work. 
 
\section{Acknowledgments\label{sec:acknowledgments}}
We thank Tim Van Hoolst and Marie Yseboodt and an unknown referee  for
careful reviews with suggestions that greatly improved and clarified
the manuscript. SJP is grateful for the support of this work provided
by the Planetary Geology and Geophysics Program of NASA under grant
NNX08Al76G.  Without the superlative
performance of the engineers and managers of The Johns Hopkins.
University Applied Physics Laboratory (APL) during the MESSENGER mission,
the values of $J_2,\,C_{22},\,C/mR^2,\,C_m/C,\,{\rm and}\,M$
used to constrain interior models of Mercury could not have been
obtained. The MESSENGER project is supported by the NASA Discovery
Program under contracts NASW-00002 to the Carnegie Institution of
Washington and NAS-97271 to APL. 

\begin{center}{\bf Appendix A: Interior surface shapes}\end{center}

We consider a model with three homogeneous layers, mantle--crust,
fluid outer core, and solid inner core. Using the generic expression
for the moment of inertia of a uniform sphere, $2mR^2/5=8\pi\rho
R^5/15$ ($R=$ radius, $m=$ mass, $\rho=$ density), we can write
expressions for $C/mR^2$ and $C_m/C$ ($C$ is principal moment of inertia
about spin axis, and $C_m$ is that of the mantle plus crust) derived from
observables in terms of $R,\,R_f,\,R_{s},\,\rho_m,\,\rho_f,\,{\rm
and}\,\rho_s$, where the subscripts 
$m,f,\,{\rm and}\,s$ refer to mantle, fluid outer core, and solid
inner core, respectively. A third equation
results from the total mass in terms of 
the same variables. With $R_f/R\rightarrow R_f,\,R_s/R\rightarrow
R_s,\,\rho_m/\bar\rho\rightarrow \rho_m,\,\rho_f/\bar\rho\rightarrow
\rho_f,\,{\rm and}\,\rho_s/\bar\rho\rightarrow \rho_s$ as normalized
variables, where $\bar\rho=$ the mean density of Mercury,

\begin{eqnarray}
R_s^3\rho_s+(R_f^3-R_s^3)\rho_f+(1-R_f^3)\rho_m&=& 1 \nonumber\\
\frac{2}{5}\left[R_s^5\rho_s+(R_f^5-R_s^5)\rho_f
+(1-R_f^5)\rho_m\right]&=&\frac{C}{mR^2}\nonumber\\
 \left[R_s^5\rho_s+
(R_f^5-R_s^5)\rho_f+(1-R_f^5)\rho_m\right] 
\frac{C_m}{C}&=&(1-R_f^5)\rho_m \label{eq:rhorcmb} 
\end{eqnarray}
Eqs. (\ref{eq:rhorcmb}) are three equations in the five unknowns
$\rho_m,\,\rho_f,\,\rho_s,\,R_f,\,{\rm and}\,R_s$. If we specify $R_s$ and
$\rho_s$, the radius and density of the solid inner core, the equations
can be solved for the remaining unknowns. 

To determine the gravitational distortion of the CMB and the inner
core radius, we need expressions for the harmonic coefficients $J_2$
and $C_{22}$ in terms of the ellipticities of the surfaces. For
uniform ellipsoids with axes $a>b>c$, the principal moments of inertia
$A<B<C$ are given by $A=M(b^2+c^2)/5,\,B=M(a^2+c^2)/5,\,C=
M(a^2+b^2)/5$ with $M=4\pi abc\rho/3$. Then $B-A=M(a^2-b^2)/5=
M(a-b)(a+b)a/5a\approx 8\pi\rho\xi r_0^5/15$ to first order in the
equatorial ellipticity $\xi=(a-b)/r_0$, where $r_0$ is the mean radius
of the ellipsoid. Similarly $C-A=8\pi\rho\epsilon_a r_0^5/15$ and
$C-B=8\pi\rho\epsilon_b r_0^5/15$ to first order in the polar ellipticities
$\epsilon_a=(a-c)/r_0$ and $\epsilon_b=(b-c)/r_0$. For a three-layer model
with each layer homogeneous,
\begin{eqnarray}
C_{22}=\frac{B-A}{4mR^2}&=&\sum_{i=1}^3\frac{\rho_i}{10}
\left(\xi_iR_i^5-\xi_{i-1}R_{i-1}^5\right)\nonumber\\
&=&\frac{1}{10}\left[\rho_m\xi_m+(\rho_f-\rho_m)\xi_fR_f^5+
(\rho_s-\rho_f)\xi_sR_s^5\right]\label{eq:c22}\\    
J_2=\frac{1}{mR^2}\left(\frac{C-A}{2}+\frac{C-B}{2}\right)&=&
\sum_{i=1}^3\frac{2\rho_i}{5}\left(\epsilon_iR_i^5-\epsilon_{i-1}R_{i-1}^5
\right)\nonumber\\
&=&\frac{2}{5}\left[\rho_m\epsilon_m+(\rho_f-\rho_m)\epsilon_fR_f^5+
(\rho_s-\rho_f)\epsilon_sR_s^5\right],\label{eq:j2}
\end{eqnarray}
where $\epsilon=(\epsilon_a+\epsilon_b)/2$ is a mean polar ellipticity and
the variables are normalized as in Eqs. (\ref{eq:rhorcmb}). If we
choose $\rho_s$ and $R_s$, and use Eqs. (\ref{eq:rhorcmb}) to
determine $\rho_m,\,\rho_f$, and $R_f$ from the observables
$M,\,C/mR^2,\,{\rm and}\,C_m/C$, the additional observables $J_2$ and
$C_{22}$ in Eqs. (\ref{eq:c22}) and (\ref{eq:j2}) give us two
equations with the 
$\xi$ and $\epsilon$ values as unknowns. To obtain additional equations in
these unknowns we assume that Mercury is hydrostatic so that
boundaries between layers of different densities are equipotential
surfaces, which we now determine. 

For our homogeneous layers we assume a surface of the following form.
\begin{equation}
r=r_0\left[1-\frac{2\epsilon}{3}P_{20}(\cos{\theta})+\frac{\xi}{6}
P_{22}(\cos{\theta})\cos{2\phi}\right],\label{eq:surface}
\end{equation}
where $r_0$ is a mean radius, and $P_{20}\,{\rm and}\,P_{22}$ are Legendre
functions. This expression leads to a surface mass distribution 
\begin{equation}
\sigma(\theta,\phi)=\sigma_0-\frac{2}{3}\epsilon r_0\rho
P_{20}(\cos{\theta})+\frac{1}{6}\xi r_0\rho
P_{22}(\cos{\theta})\cos{2\phi}, \label{eq:surfacemass} 
\end{equation}  
where $\sigma_0$, a mean surface mass density, is typically set to
zero. The potential at a point $r(\theta,\phi)>r_0$ of a surface mass
element $dm=\sigma(\theta,\phi)dA$, where $dA=r'^2\sin{\theta}d\theta d\phi$
is the element of surface area, is
\begin{eqnarray}
dV&=&-\frac{Gdm}{|\bf r-r'|}=-\frac{G\sigma(\theta',\phi')dA} {(r'^2
+r^2-2rr'\cos{S})^{1/2}}\nonumber\\ 
 &=&-\frac{G\sigma(\theta',\phi')dA}{r}\sum_{l=0}^\infty\left(\frac{r'} 
{r}\right)^l P_l(\cos{S})\nonumber\\  
&=&-\frac{G\sigma(\theta',\phi')r'^2\sin{\theta'}d\theta'd\phi'}{r}
\nonumber\\  
&&\times\sum_{l=0}^\infty\sum_{m=0}^l(2-\delta_{om})\frac{(l-m)!}{(l+m)!}
\left(\frac{r'}{r}\right)^lP_{lm}(\cos{\theta})P_{lm}(\cos{\theta'})
\cos{m(\phi-\phi')},\label{eq:dV}
\end{eqnarray}
where $S$ is the angle between $\bf r'$ and $\bf r$ and where the
primed coordinates refer to the source point $dm$ 
$(r'=r'(\theta',\phi'))$, and the unprimed coordinates $(r=r(\theta,\phi))$
to the field point where the potential is
evaluated. In Eq. (\ref{eq:dV}) $P_l$ are Legendre polynomials,
$P_{lm}$ are Legendre functions, and $\delta_{0m}$ is the Kronecker delta.
Substitution of Eq.(\ref{eq:surfacemass}) into the last of
Eqs. (\ref{eq:dV}) and integration over the surface yields terms only
for $(l,m)=(0,0),\,(2,0),\,{\rm and}\,(2,2)$ because of the
orthogonality of the Legendre functions.  Replacing $r'$ by $r_0$, we
find the potential external to the given surface distribution of mass,
\begin{equation}
V_{ext}=-\frac{4\pi G\sigma_0r'^2}{r}+\frac{8\pi}{15}\frac{G\rho
r_0^5\epsilon}{r^3}\left(\frac{3}{2}\cos^2{\theta}-\frac{1}{2}\right)
-\frac{6\pi}{15}\frac{G\rho r_0^5\xi}{r^3}\sin^2{\theta}\cos{2\phi}.
\label{eq:vext}   
\end{equation}
If $r<r_0$, the expansion is 
\begin{equation}
dV=\frac{-G\sigma(\theta',\phi')dA}{r_0}\sum_{l=0}^\infty\left(\frac{r}{r_0}
\right)^lP_{l}(\cos{S}).\label{eq:dvint}
\end{equation}
In Eq. (\ref{eq:dvint}), $r^2/r_0^3$ replaces $r_0^2/r^3$ in $V_{ext}$
for the $l=2$ terms.  For the $l=0$ term, $r_0$ in the denominator
and numerator cancels, so the potential interior to the surface
distribution of mass becomes
\begin{equation}
V_{int}=-4\pi G\sigma_0r_0+\frac{8\pi}{15}Gr^2\rho\epsilon
\left(\frac{3}{2}\cos^2{\theta}-\frac{1}{2}\right)-
\frac{6\pi}{15}Gr^2\rho\xi\sin^2{\theta}\cos{2\phi}.\label{eq:vint}  
\end{equation}
The centrifugal potential is given by 
\begin{equation}
V_{rot}=\frac{\dot\psi^2r^2}{3}[P_{20}(\cos{\theta})-1],\label{eq:centrif}
\end{equation}
where $\dot\psi=3n/2$ is the rotation rate. The lowest-order
external potential due to the Sun is averaged around the orbit to
remove the variable part as Mercury rotates and changes its distance.
\begin{equation}
\langle V_\sun\rangle=-n^2\left[-\frac{1}{2}\frac{r^2}{(1-e^2)^{3/2}}
P_{20}(\cos{\theta}) +r^2\left(\frac{7e}{2}-\frac{123e^3}{16}+\cdots
\right)P_{22}(\cos{\theta})\cos{2\phi}\right].\label{eq:sunpotential}
\end{equation}
Comparison of Eqs. (\ref{eq:centrif}) and (\ref{eq:sunpotential})
shows that they are comparable in magnitude. We indicate below that the
centrifugal potential causes a change in the ellipticities of the
boundaries of the surfaces only in the third significant figure, so
both the centrifugal and the averaged solar potential will be omitted
when calculating the ellipticities although they are displayed in the
equations.

In the following we can ignore central and constant terms.
The potential at the CMB is the sum of internal and external
potentials of the various layers, all evaluated at $r=R_f$. We can
consider the $P_{20}$ and $P_{22}$ terms separately. Positive
contributions to the magnitude of the CMB potential from the $P_{22}$
terms come from the internal potential for the outer surface with density
$\rho_m$ and ellipticity $\xi_m$, from the 
internal potential at the CMB with density $\rho_f$ and ellipticity
$\xi_f$, and from the external potential from the inner core boundary
(ICB) with density $\rho_s$ and ellipticity $\xi_s$. Negative 
contributions at the CMB come from material removed from the mantle (from
the protrusion of the ellipsoidal outer core)
with density $\rho_m$ and ellipticity $\xi_f$ and from material removed
from the fluid outer core at the ICB with density $\rho_f$ and ellipticity
$\xi_s$. If this is to be an equipotential surface it must equal
$-gh$ where $g$ is the local acceleration of gravity and $h$ is given
by the $P_{22}$ term in Eq. (\ref{eq:surface}). Then
\begin{eqnarray*}
\left\{-\frac{n^2R_f^2}{4}\left(\frac{7e}{2}-\cdots\right)
+\frac{2\pi G}{15}\left[\rho_m\xi_m R_f^2+(\rho_f-\rho_m) 
\xi_fR_f^2+(\rho_s-\rho_f)\xi_s\frac{R_s^5}{R_f^3}\right]\right\}
P_{22}(\cos{\theta})\cos{2\phi}&=&\\
\frac{4\pi
G}{3}\left[\rho_fR_f+(\rho_s-\rho_f)\frac{R_s^3}{R_f^2}\right]
\frac{\xi_fR_f}{6}P_{22}(\cos{\theta})\cos{2\phi},&&
\end{eqnarray*} 
where we have added the averaged potential due to the Sun.  
The bracketed term on the right side of the above equation is the
acceleration of gravity at the CMB, and the variables are now
dimensioned. With cancellations
\begin{equation}
-\frac{n^2R_f^2}{4}\left(\frac{7e}{2}\cdots\right)+ \rho_m\xi_m
R_f^2+(\rho_f-\rho_m) 
\xi_fR_f^2+(\rho_s-\rho_f)\xi_s\frac{R_s^5}{R_f^3}=
\frac{5}{3}\left[\rho_fR_f^2+(\rho_s-\rho_f)\frac{R_s^3}{R_f}\right]
\xi_f \label{eq:p22cmb}
\end{equation}
A similar procedure leads to an expression for the $\epsilon$s from the
equipotential surface at the CMB, where $h$ is now the $P_{20}$ term
in Eq. (\ref{eq:surface}), and where we have added the non--radial part
of the centrifugal potential and the averaged potential from the Sun.
\begin{equation}
\frac{\dot\psi_m^2R_f^2}{3}+\frac{n^2r^2}{2(1-e^2)^{3/2}}+
\rho_m\epsilon_mR_f^2+(\rho_f-\rho_m)\epsilon_fR_f^2+(\rho_s-\rho_f) 
\epsilon_s\frac{R_s^2}{R_f^3}=\frac{5}{3}\left[\rho_fR_f^2+(\rho_s-
\rho_f)\frac{R_s^3}{R_f}\right]\epsilon_f \label{eq:p20cmb}
\end{equation}   

For the ICB only inner potentials are involved, all evaluated at
$r=R_s$. Again setting the potential at the ICB equal to the
local $-gh$, we find for the $\xi$s and $\epsilon$s
\begin{eqnarray}
-\frac{n^2R_s^2}{4}\left(\frac{7e}{2}+\cdots\right)+ \rho_m\xi_m
+(\rho_f-\rho_m)\xi_f+(\rho_s-\rho_f)\xi_s&=& 
\frac{5}{3}\rho_s\xi_s\label{eq:p22icb}\\
\frac{\dot\psi_s^2R_s^2}{3}+\rho_m\epsilon_m+(\rho_f-\rho_m)\epsilon_f+(\rho_s-\rho_f)\epsilon_s&=& 
\frac{5}{3}\rho_s\epsilon_s.\label{eq:p20icb}
\end{eqnarray}
Eqs. (\ref{eq:c22}), (\ref{eq:p22cmb}), and (\ref{eq:p22icb}) are three
equations in the three unknowns $\xi_i$, and Eqs. (\ref{eq:j2}),
(\ref{eq:p20cmb}), and (\ref{eq:p20icb}) are three equations in the
three unknowns $\epsilon_i$. Table \ref{tab:alphabeta} in the main text
gives the solutions for $\rho_s=8\,{\rm g/cm^3},\,R_s=0.6R$ and for $R_s=0$
(no solid inner core). Because of Mercury's slow rotation, the
centrifugal contribution to the potentials makes a change only in the
third significant decimal place in the solutions for the $\epsilon_i$,
so it and the comparable contribution from the averaged solar
potential are neglected in these solutions. 

\begin{center}{\bf Appendix B: Potential contribution to
pressure torque}\end{center}

The contribution to the pressure torque from the external potential
$\Phi$ in Eqs. (\ref{eq:presstorque}) and (\ref{eq:delp}) is
\begin{equation}
{\bf\Gamma}_{P\sun}=-\iint_S{\bf r'}\times{\bf n}\rho_f\Phi dS,\label{eq:rxnphi}
\end{equation}
where ${\bf r'}(\theta',\phi')$ is a radius vector to a point on the CMB
measured from the center of Mercury, {\bf n} is the surface normal and
\begin{equation}
\Phi=-\frac{Gm_{\sun}}{r}\sum_{l=0}^\infty\sum_{m=0}^l\left(\frac{r'}{r}
\right)^l(2-\delta_{0m})\frac{(l-m)!}{(l+m)!}P_{lm}(\cos{\theta})
P_{lm}(\cos{\theta'})\cos{m(\phi-\phi')},\label{eq:phisun}
\end{equation}
is the potential of the Sun interior to Mercury, with $r,\theta,\phi$
the spherical coordinates of the Sun in Mercury's principal axis
system.   With the surface of the CMB represented by $r'=
R_f(1-(2\epsilon_f/3)P_{20}(\cos{\theta'})+(\xi_f/6)P_{22}(\cos{\theta'})
\cos{2\phi'}$ (Eq.(\ref{eq:surface})), we can form the function
\begin{equation}
H=r'-R_f[1-(2\epsilon_f/3)P_{20}(\cos{\theta'})+(\xi_f/6)P_{22}
(\cos{\theta'})\cos{2\phi'}],\label{eq:hrthetaphi}
\end{equation}
so that
\begin{equation}
{\bf n}=\frac{\nabla H}{|\nabla H|}={\bf{\hat e}}_r-\cos{\theta'}
\sin{\theta'}\left(2\epsilon_f+\xi_f\cos{2\phi}\right){\bf{\hat
e}}_{\theta'}+\xi_f\sin{\theta'} 
\sin{2\phi'}{\bf{\hat e}}_{\phi'},\label{eq:n}
\end{equation}
where $R_f/r'=1$ to first
order in $\epsilon$ or $\xi$ ($|\nabla H|$ is O$(1+\epsilon^2\,{\rm
or}\,\xi^2)$) and 
\begin{equation}
{\bf r'}\times{\bf n}=-R_f[\cos{\theta'}\sin{\theta'}\left(2\epsilon_f
+\xi_f\cos{2\phi}\right){\bf{\hat e}}_{\phi'}+\xi_f\sin{\theta'}
\sin{2\phi'}{\bf{\hat e}}_{\theta'}].\label{eq:rxn}
\end{equation}
Substitution of Eqs. (\ref{eq:rxn}) and (\ref{eq:phisun}) into
Eq. (\ref{eq:rxnphi}) yields the pressure torque on the CMB from the
solar potential.  For the integration over the surface, we convert the
unit vectors in spherical coordinates to unit vectors in Cartesian
coordinates with ${\bf{\hat 
e}_\phi'}=-\sin{\phi'}{\bf i}+\cos{\phi'}{\bf j}$ and ${\bf{\hat
e}}_{\theta'}=\cos{\theta'}\cos{\phi'}{\bf
i}+\cos{\theta'}\sin{\phi'}{\bf j}-\sin{\theta'}{\bf k}$. All the terms
in the ${\bf i}$ and ${\bf j}$ components of the integrand contain
the product $\cos{\theta'}\sin{\theta'}$, which select only the
$P_{21}(\cos{\theta'}) =-3\sin{\theta'}\cos{\theta'}$ term in $\Phi$,
and the ${\bf k}$ component contains $\sin^2{\theta'}$, which selects
only the $P_{22}(\cos{\theta'})$ term in $\Phi$ from the orthogonality of
the Legendre functions. Integration yields
\begin{eqnarray}
{\bf\Gamma}_{P\sun}&=&\frac{4\pi}{15}\frac{Gm_\sun}{r^3}R_f^5\rho_f
\big[(-2\epsilon_f+\xi_f)P_{21}(\cos{\theta})\sin{\phi}{\bf
i}\nonumber\\ 
&&+(2\epsilon_f+\xi_f)P_{21}(\cos{\theta})\cos{\phi}{\bf j}+
\xi_fP_{22}(\cos{\theta})\sin{2\phi}{\bf k}\big].\label{eq:sunpresstorque}
\end{eqnarray}
Recall from  Appendix A that $B'-A'=8\pi\rho\xi
R^5/15$, to first order in $\xi$, is the moment of inertia difference
of a thin layer of material outside the largest sphere that would fit
inside the ellipsoid. Similarly $C'-A'$ and $C'-B'$ are the same
expressions with $\xi$ replaced by $\epsilon_a$ and $\epsilon_b$,
respectively, and $\epsilon=(\epsilon_a+\epsilon_b)/2$.  Then we can rewrite
the components of Eq. (\ref{eq:sunpresstorque}) as 
\begin{eqnarray}
({\bf\Gamma}_{P\sun})_x&=&\frac{Gmm_{\s\sun}R^2}{r^5}({\bf r\cdot
k})({\bf r\cdot j})(3J'_2-6C'_{22}),\nonumber\\ 
({\bf\Gamma}_{P\sun})_y&=&-\frac{Gmm_{\s\sun}R^2}{r^5}({\bf r\cdot
k})({\bf r\cdot i})(3J'_2+6C'_{22}),\nonumber\\ 
({\bf\Gamma}_{P\sun})_z&=&12\frac{Gmm_{\s\sun}R^2}{r^5}({\bf r\cdot
i})({\bf r\cdot j})C'_{22},\label{eq:sunpresstorque2}
\end{eqnarray}
where $J'_2=[(C'-A')/2+(C'-B')/2]/mR^2$ and
$C'_{22}=(B'-A')/4mR^2$. The right sides of these equations are
identical to those of Eq. (\ref{eq:torque1}) with $J_2\,{\rm and}\,C_{22}$
replaced by $J'_2\,{\rm and}\,C'_{22}$.  The pressure torque due to the solar
potential simply adds to the gravitational torque on the mantle as if
the thin layer of fluid of density $\rho_f$ were added to the
mantle. These additions are the contributions of the core to $J_2$ and
$C_{22}$. So if the inner core contributions are small or otherwise
neglected, the gravitational torque on the mantle is that appropriate
to the total $J_2$ and $C_{22}$ and not just the values appropriate to
the mantle--crust alone.

\newpage
\parindent=0pt
\begin{center}References\end{center}
Anderson, B.J., Johnson, C.L., Korth, H, Purucker, M.E., Winslow, R.M.,
Slavin, J.A., Solomon, S.C., McNutt, R.L.Jr., Raines, J.M., Zurbuchen, T.H.,
2011. The global magnetic field of Mercury from MESSENGER orbital
observation.  Science  333, 1859--1862.

Anderson, B.J., Johnson, C.L., Korth, H., Winslow, R.M., Borovsky,
J.E., Purucker, M.E., Slavin, J.A., Solomon, S.C., Zuber, M.T.,
McNutt, R.L.Jr., 2012. Low-degree structure in Mercury's magnetic
field. J. Geophys. Res. 117, E00L12, doi: 10.129/2012JE001459.

Buffett, B.A., 1992. Constraints on magnetic energy and mantle conductivity
from forced nutations of the Earth. J. Geophys. Res. 97,
19581--19597. 

Busse, F.H., 1968. Steady flow in a precessing spheroidal shell. 
J. Fluid. Mech. 33, 739--751.

Castillo-Rogez, J.C., Efroimsky, M., Lainey, V., 2011. The tidal history
of Iapetus: Spin dynamics in the light of a refined dissipation model.
J. Geophys. Res. 116, E09008, doi: 10.1029/2010JE003664.

Clauser, C., Huenges, E., 1995. Thermal conductivity of rocks and
minerals, In: Ahrens, T.J. Ed., Rock Physics and Phase Relations, A
Handbook of Physical Constants, AGU Reference Shelf 3, Washington, DC,
pp. 105--126.  

Colombo, G., 1966. Cassini's second and third laws. Astron. J.
71, 891--896.

Colombo, G., Shapiro, I.I., 1966. The rotation of the planet
Mercury. Astrophys. J. 145, 296--307. 

Correia, A.C.M., Laskar, J., 2004. Mercury's capture into the 3/2
spin--orbit resonance as a result of its chaotic dynamics. Nature
429, 848--850.

Correia, A.C.M., Laskar J., 2009. Mercury's capture into the 3/2
spin--orbit resonance including the effect of core--mantle interaction.
Icarus 201, 1--11.

de Wijs, G.A., Kresse, G., Vo\v{c}adlo, L., Dobson, D., Alf\`e, D.,
Gillan, M.J., Price, G.D., 1998. The viscosity of liquid iron at the
physical conditions of the Earth's core. Nature 392, 805--807. 

Goldreich, P., Peale, S.J., 1966. Spin--orbit coupling in the solar
system.  Astron. J. 71, 425--438.

Greenspan, H.P., Howard, L.N., 1963. On a time-dependent motion of a
rotating fluid. J. Fluid Mech. 17, 385--404.

Hauck, S.A.II, Margot, J.L., Solomon, S.C., Phillips, R.J., Johnson, C.L.,
Lemoine, F.G., Mazarico, E.,  McCoy, T.J., Padovan, S., Peale, S.J.,
Perry, M.A., Smith, D.E., Zuber M.T., 2013. The curious case of Mercury's
internal structure. J. Geophys. Res. Planets 118, 1204--1220.

Hut, P., 1981. Tidal evolution in close binary systems. 
Astron. Astrophys. 99, 126--140. 

Margot, J.L., Peale, S.J., Jurgens, R.F., Slade, M.A. Holin, I.V., 2007.
Large longitude libration of Mercury reveals a molten core. Science
316, 710--714.

Margot, J.L., Peale, S.J., Solomon, S.C., Hauck, S.A.II,  Ghigo, F.D.,
Jurgens, R.F., Yseboodt, M., Giorgini, J.D., Padovan, S., Campbell, D.B.,
2012. Mercury's moment of inertia from spin and gravity data, 
J. Geophys. Res.  117, E00L09, doi: 10.1029/2012JE004161.

Melchior, P., 1986.  The Physics of the Earth's Core, Pergamon
Press, Oxford, pp. 218--222.   

Michel, N.C., Hauck, S.A., Solomon, S.C., Phillips, R.J., Roberts,
J.H., Zuber, M.G. (2013) Thermal evolution of Mercury as constrained
by MESSENGER observations, {\it J. Geophys. Res. Planets} {\bf 118},
1033-1044. 

Mignard, F., 1979. The evolution of the lunar orbit revisited, I.
Moon Planets 20, 301--315.

Mignard, F., 1980 The evolution of the lunar orbit revisited, II.
Moon Planets 23, 185--201.

Mignard, F., 1981. The evolution of the lunar orbit revisited, III.
Moon Planets 24, 189--207.

Peale, S.J., 1969. Generalized Cassini's laws.  Astron. J. 
74, 483--489.

Peale, S.J., 1974. Possible histories of the obliquity of Mercury.
Astron. J. 79, 722--744.

Peale, S.J., 2005. The free precession and libration of Mercury.
Icarus 178, 4--18.

Peale, S.J., 2006. The proximity of Mercury's spin to Cassini state 1
from adiabatic invariance.  Icarus 181, 338--347. 

Peale, S.J., 2007 The origin of the natural satellites. In Schubert,
G., Spohn, T. (Eds.) Planets and Moons, Treatise on Geophysics, Volumn
10: Moons, Elsevier, pp. 465--508. 

Peplowski, P.N., Evans, L.G. Hauck II, S.A., McCoy, T.J., Boynton,
W.V., Gillis-Davis, J.J. Ebel, D.S., Goldsten, J.O., Hamara, D.K.,
Lawrence, D.J., McNutt Jr., R.L., Nittler, L.R., Solomon, S.C., Rhodes,
E.A., Sprague, A.L., Starr, R.D., Stockstill-Cahill,
K.R. 2011. Radioactive elements on Mercury's surface from MESSENGER:
Implications for the Planet's formation and evolution. Science 333,
850-852. 

Pettengill, G.H., Dyce, R.B., 1965. A radar determination of the
rotation of the planet Mercury. Nature 206, 1240.

Poincar\'e, H., 1910. Sur la pr\'ecession des corps d\'eformables. 
Bull. Astron.  27, 321--367.  

Popov, P.A., Fedorov, P.P., Kuznetsov, S.V., 2013. Thermal
conductivity of ${\rm FeS_2}$ pyrite crystals in the temperature range
50-300 K. Crystal. Rep. 58, 319--321.
 
Roberts, P.H., Stewartson, K., 1965. On the motion of a liquid in a
spheroidal cavity of a precessing rigid body.  J. Geophys. Res.
 67, 279--288.

Smith, D.E., Zuber, M.T., Phillips, R.J., Solomon, S.C., Hauck, S.A.II.,
Lemoine, F.G., Mazarico, E.,  Neuman, G.A., Peale, S.J., Margot, J.L.,
Johnson, C.L., Torrence, M.H., Perry, M.E.,  Rowlands, D.G., Goossens,
S., Head, J.W., Taylor, A.H., 2012. Gravity field and internal structure of
Mercury from MESSENGER. Science 336, 214--217.
 
Smylie, D.E., Brazhkin, V., Palmer, A., 2009. Direct observations of
the viscosity of Earth's outer core and extrapolation of measurements
of the viscosity of liquid iron. Physics-Uspekhi 52, 79--92.

Stewartson, K., Roberts P.H., 1963. On the motion of a liquid in a
spheroidal cavity of a precessing rigid body.  J. Fluid Mech.
17, 1--20.

Tosi, N., Grott, M., Plesa, A-C. Breuer, D. 2013, A Thermo-chemical
evolution of Mercury's interior, {\it J. Geophys. Res. Planets}, {\bf
118}, doi: 10.1002/jgre.20168. 

Turcotte, D.L., Schubert, G. 2002 Geodynamics (2nd ed.), Cambridge
U. Press, Cambridge, UK, p. 137.

Van Hoolst, T., Jacobs, C. 2003 Mercury's tides and interior
structure, J. Geophys. Res. 108 (E11), 5121, doi:10.1029/2003JE002126.

Van Hoolst, T. A. Rivoldini, R.M. Baland, M. Yseboodt, (2012)
The effect of tides and an inner core on the forced 
longitudinal librations of Mercury, Earth Planet. Sci.
Let., {\bf 333-334}, 83-90.

Vo\v{c}adlo, L., 2007. Core viscosity. In: Gubbins, D., Herro-Bervera,
E. (Eds.), Encyclopedia of Geomagnetism and Paleomagnetism, Springer,
pp. 104--106.
\end{document}